\documentclass{vldb}

\usepackage{endnotes}
\usepackage{graphicx} 
\usepackage{booktabs} 
\usepackage{xspace} 
\usepackage{balance} 
\usepackage{subfigure} 
\usepackage{xcolor} 
\usepackage[colorlinks]{hyperref} 
\usepackage{enumitem} 
\usepackage{amsmath} 

\definecolor{darkred}{rgb}{0.5,0,0}
\definecolor{darkgreen}{rgb}{0,0.5,0}
\definecolor{darkblue}{rgb}{0,0,0.5}
\hypersetup{colorlinks,linkcolor=darkblue,filecolor=darkgreen,urlcolor=darkred,citecolor=darkblue}

\newcommand{\codename}{STAR\xspace}
\newcommand{\twopl}{Dist.~S2PL\xspace}
\newcommand{\occ}{Dist.~OCC\xspace}
\newcommand{\pb}{PB.~OCC\xspace}

\newcommand{\tpcc}{\mbox{TPC-C}\xspace}

\vldbTitle{\codename: Scaling Transactions through Asymmetric Replication}
\vldbAuthors{Yi Lu, Xiangyao Yu and Samuel Madden}
\vldbDOI{https://doi.org/10.14778/3342263.3342270}
\vldbPageFrom{1316}
\vldbPageTo{1329}
\vldbVolume{12}
\vldbNumber{11}
\vldbYear{2019}

\begin{document}

\title{\codename: Scaling Transactions through Asymmetric Replication}

\numberofauthors{3}
\author{
\alignauthor
Yi Lu\\
\affaddr{MIT CSAIL}\\
\email{yilu@csail.mit.edu}
\alignauthor
Xiangyao Yu\\
\affaddr{MIT CSAIL}\\
\email{yxy@csail.mit.edu}
\alignauthor
Samuel Madden\\
\affaddr{MIT CSAIL}\\
\email{madden@csail.mit.edu}
}

\maketitle

\begin{abstract}

In this paper, we present \codename,  a new distributed in-memory 
database with asymmetric replication. 
By employing a single-node non-partitioned architecture for some replicas and a partitioned architecture for other replicas, 
\codename is able to
efficiently run both highly partitionable workloads and workloads that involve
cross-partition transactions.  The key idea is a new {\it phase-switching} 
algorithm where the execution of single-partition and cross-partition transactions is separated.
In the partitioned phase, single-partition transactions are run on multiple machines in parallel to exploit more concurrency. 
In the single-master phase, mastership for the entire database is switched to a single designated master node, which can
execute these transactions without the use of expensive coordination protocols
like two-phase commit.  Because the master node has a full copy of the database, this phase-switching 
can be done at negligible cost. Our experiments on two popular benchmarks (YCSB
and \tpcc) show that high availability via replication can coexist with fast
serializable transaction execution in distributed in-memory databases,  with
\codename outperforming systems that employ conventional concurrency control and replication algorithms by 
up to one order of magnitude.

\end{abstract}
\section{Introduction} \label{sec:introduction}
Recent years have seen a number of in-memory transaction processing systems that 
 can run hundreds of thousands to millions of
transactions per second by leveraging multi-core parallelism~\cite{StonebrakerMAHHH07, TuZKLM13,
YuPSD16}.   These systems can be broadly classified into i) partitioning-based
systems, e.g., H-Store~\cite{StonebrakerMAHHH07} which partitions data onto 
different cores or machines, and ii) non-partitioned systems that try to minimize the overheads associated 
with concurrency control in a single-server, 
multi-core setting by avoiding locks and contention whenever possible~\cite{NarulaCKM14, NeumannMK15, TuZKLM13,YuPSD16}, 
while allowing any transaction to run on any core/processor. 

\begin{figure}[!t]
    \centering
    \includegraphics[width=0.9\columnwidth]{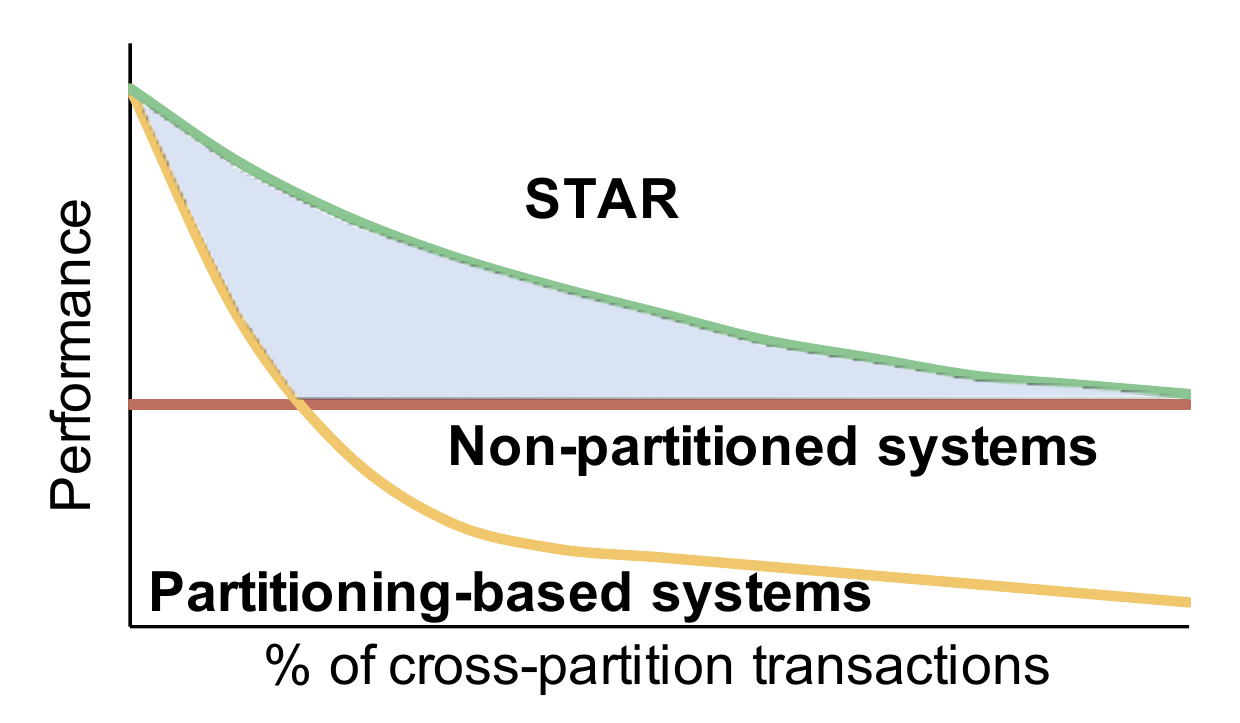}
    \vspace{-5mm}
    \caption{Partitioning-based systems vs. Non-partitioned systems} \label{fig:cartoon}
    \vspace{-6mm}
\end{figure}

As shown in Figure~\ref{fig:cartoon}, partitioning-based systems work well when workloads have few cross-partition transactions, 
because they are able to avoid the use of any synchronization and thus scale out to multiple machines. 
However, these systems suffer when transactions need to access more than one partition, especially in a distributed setting where expensive
 protocols like two-phase commit are required to coordinate these cross-partition transactions.  
In contrast, non-partitioned approaches provide somewhat lower performance on highly-partitionable workloads due to their inability to scale out, 
but are not sensitive to how partitionable the workload is.

In this paper, we propose a new transaction processing system, \codename, that is able to achieve the best of both worlds.  
We start with the observation that most modern transactional systems will keep several replicas of data  for high availability purposes.
In \codename, we ensure that at least one of these replicas is {\it complete}, i.e., it stores all records on a single machine, as in the recent non-partitioned approaches.  
We also ensure that at least one of the replicas is partitioned across several processors or machines, as in partitioned schemes.  
The system runs in two phases, using a novel {\it phase-switching} protocol:  
in the {\it partitioned} phase, transactions that can be executed completely on a single partition are mastered at one of the partial replicas storing that partition, 
and replicated to all other replicas to ensure fault tolerance and consistency.  
Cross-partition transactions are executed during the {\it single-master}  phase, during which mastership for all records is switched to one of the complete replicas, which runs the transactions to completion and replicates their results to the other replicas.   
Because this node already has a copy of every record, changing
the master for a record ({\it re-mastering}) can be done without transferring any data, 
ensuring lightweight phase switching. 
Furthermore, because the node has a complete replica, transactions can be coordinated within a single machine without a slow commit protocol like two-phase commit.   
In this way, cross-partition transactions do not incur commit overheads as in typical partitioned approaches (because they all run on the single master), 
and can be executed using a lightweight concurrency control protocol like Silo~\cite{TuZKLM13}. 
Meanwhile, single-partition transactions can still be executed without any concurrency control at all, as in H-Store~\cite{StonebrakerMAHHH07}, 
and can be run on several machines in parallel to exploit more concurrency.
The latency that phase switching incurs is no larger than the typical delay used in high-performance transaction processing systems for an epoch-based group commit.

Although the primary contribution of \codename is this phase-switching protocol, to build the system, we had to explore several novel aspects of in-memory concurrency control.  In particular, prior systems like Silo~\cite{TuZKLM13} and TicToc~\cite{YuPSD16} were not designed with high-availability (replication) in mind. 
Making replication work efficiently in these systems requires some care and is amenable to a number of optimizations.
For example, \codename uses intra-phase asynchronous replication to achieve high performance. In the meantime, it ensures consistency among replicas via a replication fence when phase-switching occurs. 
In addition, with our phase-switching protocol, \codename can use a cheaper replication strategy than
that employed by replicated systems that need to replicate entire
records~\cite{ZhengTKL14}. This optimization can significantly reduce bandwidth requirements
(e.g., by up to an order of magnitude in our experiments with TPC-C.).

Our system does require a single node to hold an entire copy of the database, 
as in many other modern transactional systems~\cite{DiaconuFILMSVZ13, FaleiroA15, KimWJP16, LimKA17, TuZKLM13, WangK16, YuPSD16}. 
Cloud service providers, such as Amazon EC2~\cite{ec2} and Google Cloud~\cite{google-cloud}, now provide high memory instances with 12~TB RAM, and 24 TB instances are coming in the fall of 2019. 
Such high memory instances are sufficient to store 10 kilobytes of online state about each customer in a database with about one billion customers, which exceeds the scale of all but the most demanding transactional workloads.
In addition, on workloads with skew, a single-node in-memory database can be further extended through an anti-caching architecture~\cite{DeBrabantPTSZ13}, 
which provides 10x larger storage and competitive performance to in-memory databases.

In summary, \codename is a new distributed and replicated in-memory 
database that employs both partitioning and replication.
It encompasses a number of interesting aspects:
\begin{itemize}[noitemsep,nolistsep]
	\item By exploiting multicore parallelism and fast networks, \codename is 
		  able to provide high throughput with serializability guarantees. 
	\item It employs a \textit{phase-switching} scheme which enables \codename
	      to execute cross-partition transactions without two-phase commit while
	      preserving fault tolerance guarantees. 
	\item It uses a hybrid replication strategy to reduce the overhead of replication, 
		while providing transactional consistency and high-availability.
\end{itemize}
 
 In addition, we present a detailed evaluation of \codename that demonstrates its ability to provide adaptivity, 
 high availability, and high-performance  transaction execution. 
\codename outperforms systems that employ conventional distributed concurrency control algorithms by up to one order of magnitude on YCSB and TPC-C. 

\section{Background} \label{sec:background}

In this section, we describe how concurrency control allows database systems to execute transactions with serializability guarantees. 
We also introduce how replication is used in database systems with consistency guarantees.

\subsection{Concurrency Control Protocols} \label{ssec:transaction_protocols}

Serializability --- where the operations of multiple transactions are
interleaved to provide concurrency while ensuring that the state of the
database is equivalent to some serial ordering of the transactions --- is the
gold standard for transaction execution.
Many serializable concurrency control protocols have been proposed, starting
from early protocols that played an important role in hiding disk latency~\cite{BernsteinHG87} to modern OLTP systems that exploit
multicore parallelism~\cite{TuZKLM13, YuPSD16}, typically employing lock-based and/or optimistic techniques.

Two-Phase locking (2PL) is the most widely used classical protocol to ensure
serializability of  concurrent transactions~\cite{EswarranGLT76}.  
 2PL is considered
pessimistic since the database acquires locks on operations even when there is no conflict. By contrast, optimistic concurrency control
protocols (OCC)  avoid this by only checking conflicts at the end of a
transaction's execution~\cite{KungR81}.  OCC runs transactions in three phases:
 read, validation, and write. In the read phase,  transactions perform
read operations from the database and write operations to local copies of objects
without acquiring locks. In the validation phase,  conflict checks are done
against all concurrently committing transactions. If conflicts exist, the
transaction aborts. Otherwise, it enters the write phase and
copies its local modifications to  the database. Modern systems, like
Silo~\cite{TuZKLM13},
typically employ OCC-like techniques because they make it easier to avoid the overhead of shared locks during query execution.

In distributed database systems, cross-partition transactions involving many machines are classically coordinated
using two-phase commit (2PC)~\cite{MohanLO86} protocol to achieve fault tolerance, since machines can fail independently.
The coordinator decides to commit or abort a transaction based on decisions collected from 
workers in the prepare phase. Workers must durably remember their decisions in the prepare phase until they learn the transaction outcome. Once the decision on the coordinator is made, 2PC enters the
commit phase, and workers commit or abort the transaction based on its decision. Although 2PC does ensure serializability, the additional overhead
of multiple log messages and network round trips for each transaction can significantly reduce throughput of distributed transactions.
In \codename, we avoid two-phase commit by employing a phase-switching protocol to re-master records in distributed transactions
to a single primary, which runs transactions locally and replicates them asynchronously.

\subsection{Replication} \label{ssec:replication}

Modern database systems need to be highly available. 
When a subset of servers in a cluster fails, the system needs to quickly reconfigure itself and replace a failed server with a standby machine, such that an end user does not experience any noticeable downtime. 
High availability requires the data to be replicated across multiple machines in order to allow for fast fail-over. 

Primary/backup replication is one of the most widely used schemes. 
After the successful execution of a transaction, the primary node sends the log to all involved backup nodes. 
The log can either contains values~\cite{DB2, SQLServer,MySQL,PostgreSQL}, which are applied by the backup nodes to the replica, or operations~\cite{QinGB17}, which are re-executed by the backup nodes. 
For a distributed database, each piece of data has one primary node and one or multiple backup nodes.
A backup node for some data can also be a primary node for some other data. 

In distributed database systems, both 2PC and replication are important to ensure ACID properties. 
They are both expensive compared to a single node system, but in different ways. In particular, 
replication incurs very large data transfer but does not necessarily need expensive coordination for each transaction.

\section{\codename Architecture} \label{sec:architecture}

\begin{figure}[!t]
\centering
\includegraphics[width=0.99\columnwidth]{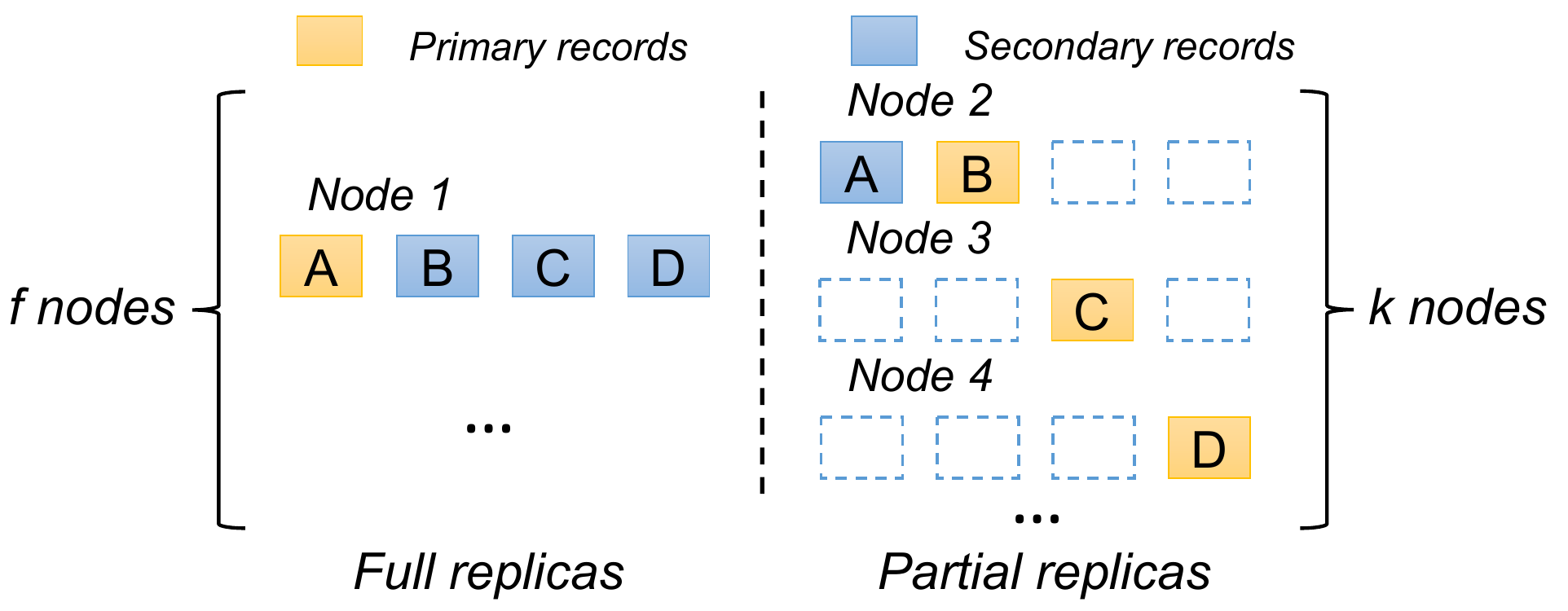}
\vspace{-6mm}
\caption{The architecture of \codename}\label{fig:architecture}
\vspace{-6mm} 
\end{figure}

\codename is a distributed and replicated in-memory database.
It separates the execution of single-partition transactions and cross-partition
transactions using a novel {\it phase-switching protocol}. 
The system dynamically switches between  {\it partitioned} and {\it single-master} phases.
In \codename, each partition is mastered by a single node.  
During the partitioned phase, queries that touch only a single partition are executed one at a time to completion 
on their partition.  
Queries that touch multiple partitions (even if those partitions are on a single machine)
are executed in the single-master phase on a single designated master node.

To support the phase-switching protocol, \codename employs asymmetric replication.
As shown in Figure~\ref{fig:architecture}, the system consists of two types of replicas, namely, (1) full replicas, and (2) partial replicas. 
Each of the $f$ nodes (left side of figure) has a full replica, which consists of all database partitions. 
Each of the $k$ nodes (right side of figure) has a partial replica, which consists of a portion of the database.
In addition, \codename requires that these $k$ partial replicas together contain at least one full copy of the database. 
During the partitioned phase, each node (whether a full or partial replica) acts as a master for some part of the database.
During the single-master phase, one of the $f$ nodes acts as the master for the whole database.
Note that writes of committed transactions are replicated at least $f+1$ times on a cluster of $f+k$ nodes.
We envision $f$ being small (e.g., 1), while $k$ can be much larger.
Having more than one full replica (i.e., $f > 1$) does not necessarily improve system performance, 
but provides higher availability when failures occur (See Section~\ref{ssec:fault_tolerance}).

\begin{figure}[!t]
\centering
\includegraphics[width=0.99\columnwidth]{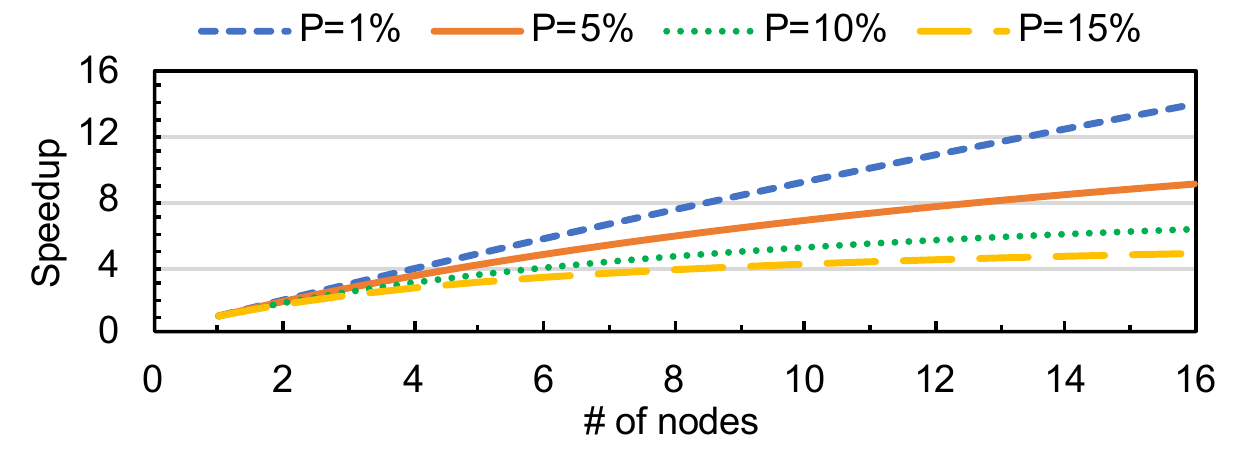}
\vspace{-6mm}
\caption{Performance speedup of asymmetric replication in \codename over single node execution;  $\mathcal{P}$ = Percentage of cross-partition transactions}\label{fig:analytical_model}
\vspace{-6mm} 
\end{figure}

There are several advantages of this phase switching approach.  
First, in the single-master phase, cross-partition transactions are run on a single master node.
As in existing non-partitioned systems,
two-phase commit is not needed as the master node runs all cross-partition transactions and propagates writes to replicas.
In contrast, transactions involving multiple nodes in existing distributed partitioning-based systems are coordinated with two-phase commit, 
which can significantly limit the throughput of distributed database systems~\cite{HardingAPS17}.
Note that transactions running in the partitioned phase also do not require two-phase commit because they all run on a single partition, on a single node.
Second, in the partitioned phase, single-partition transactions are run on multiple nodes in parallel,  
providing more concurrency, as in existing partitioning-based systems. 
This asymmetric replication approach is a good fit for workloads 
with a mix of single-partition and cross-partition transactions on a cluster of about 10 nodes.
We demonstrate this both empirically and through the use of an analytical model, 
as shown in Sections~\ref{sec:evaluation} and \ref{ssec:best_of_both_worlds}.
Figure~\ref{fig:analytical_model}
shows the speedup predicted by our model of \codename with $n$ nodes over a single node.

\codename uses a variant of Silo's OCC protocol~\cite{TuZKLM13}. 
Each record in the database maintains a transaction ID (TID) from the transaction that last updated the record.
The TID is used to determine which other transactions a committing transaction may have a conflict with. 
For example, once a transaction begins validation, it will abort if any one of the accessed records is locked by other transactions or has a different TID.
The TID is assigned to a transaction after it is successfully validated. 
There are three criteria for the TID obtained from each thread: 
(a) it must be larger than the TID of any record in the read/write set; 
(b) it must be larger than the thread's last chosen TID; 
(c) it must be in the current global epoch.
The first two criteria guarantee that the TIDs of transactions having conflicting writes are assigned in a serial-equivalent order. 
As in Silo, \codename uses a form of {\it epoch-based group commit}.
The serial order of transactions running within an epoch is not explicitly known by the system except across epoch boundaries, since anti-dependencies (i.e., write-after-read conflicts) are not tracked.  
Different from Silo, in which the global epoch is incremented every 40~ms, a phase switch in \codename naturally forms an epoch boundary. 
Therefore, the system only releases the result of a transaction when the next phase switch occurs.
At this time, the global epoch is also incremented.

At any point in time, each record is {\it mastered} on one node, with other nodes serving as secondaries. 
Transactions are only executed over primary records;  writes of committed transactions are propagated to all replicas. 
One way to achieve consistency is to replicate writes {\it synchronously} among replicas.
In other words, the primary node holds the write locks when replicating writes of a committed transaction to backup nodes. 
However, this design increases the latency of replication and impairs system performance.
In \codename, writes of committed transactions are buffered and {\it asynchronously} shipped to replicas, meaning the locks on the primary are not held during the process. 
To ensure correctness, we employ the Thomas write rule~\cite{Thomas79}:
we tag each record with the last TID that wrote it and apply a write if the TID of the write is larger than the
current TID of the record.  Because TIDs of conflicting writes are guaranteed to be assigned in the serial-equivalent
order of the writing transactions, this rule will guarantee correctness.  
In addition, \codename uses a {\it replication fence} when phase-switching occurs. 
In this way, strong consistency among replicas is ensured across the boundaries of the partitioned phase and the single-master phase.

Tables in \codename are implemented as collections of hash tables, which is typical in many in-memory
databases~\cite{WangK16, WeiSCCC15, YuPSD16}. Each table is built on top of a primary hash table and contains zero or 
more secondary hash tables as secondary indexes. To access a record, \codename probes the hash table with 
the primary key. Fields with secondary indexes can be accessed by mapping a value to the relevant 
primary key. Although \codename is built on top of hash tables, it is easily adaptable to other data 
structures such as Masstree~\cite{MaoKM12}. 
As in most high performance transactional systems~\cite{StonebrakerMAHHH07, TuZKLM13, YuPSD16},
clients send requests to \codename by calling pre-defined
\textit{stored procedures}. Parameters of a stored procedure must also be passed to \codename with the request.
Arbitrary logic (e.g., read/write operations) is supported in stored procedures, which are implemented in C++.

\codename is \textit{serializable} by default but can also support 
\textit{read committed} and \textit{snapshot isolation}. 
A transaction runs under \textit{read committed} by skipping read validation on commit, 
since \codename uses OCC and uncommitted data never occurs in the database. 
\codename can provide snapshot isolation by retaining additional versions for records~\cite{TuZKLM13}.

\section{The Phase Switching Algorithm} \label{sec:the_phase_switching_algorithm}

\begin{figure*}[!t]
    \begin{minipage}{0.48\linewidth}
    \centering
    \includegraphics[width=0.99\columnwidth]{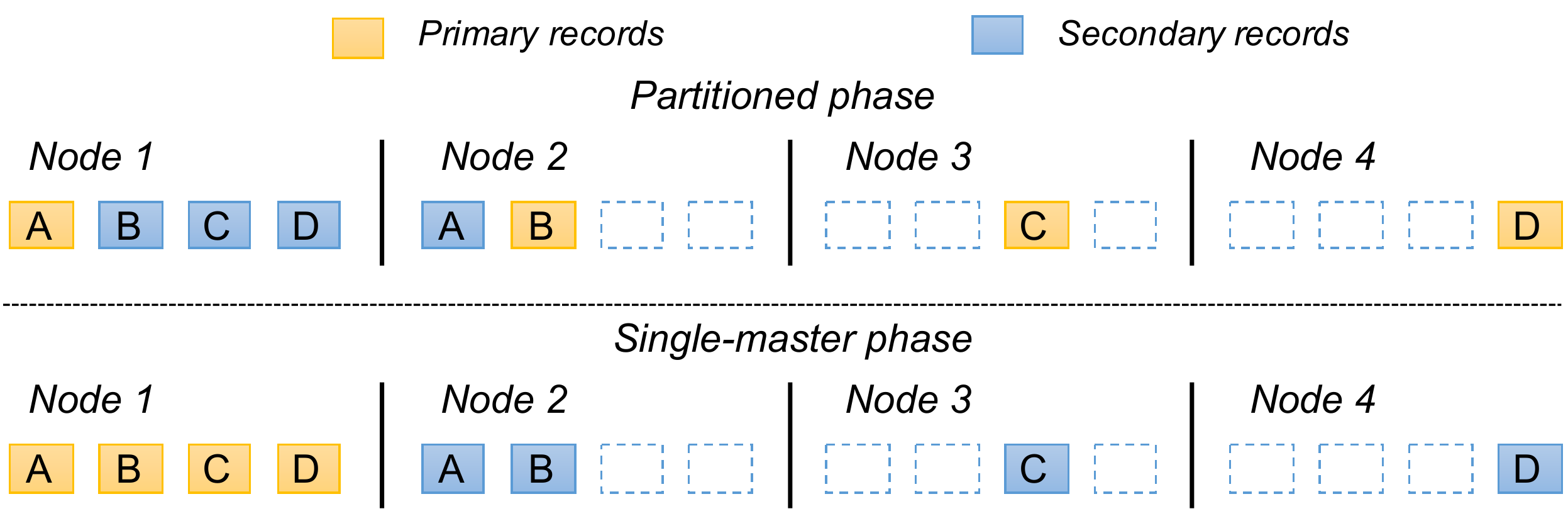}
    \vspace{-4mm}
    \caption{Illustrating the two execution phases} \label{fig:phase_switching}
    \vspace{-4mm}
    \end{minipage}
    \hspace{0.02\linewidth}
    \begin{minipage}{0.48\linewidth}
    \centering
    \includegraphics[width=0.99\columnwidth]{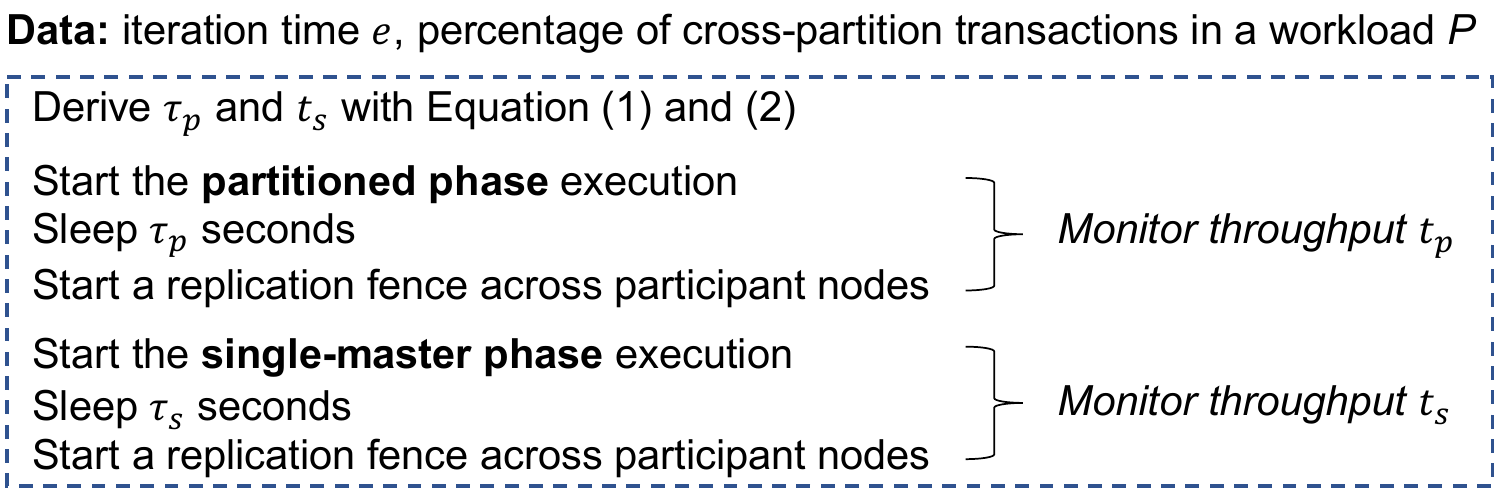}
    \vspace{-4mm}
    \caption{Phase transitions in \codename}\label{alg:phase_switch_code}
    \vspace{-4mm}   
    \end{minipage}
\end{figure*}

We now describe the phase switching algorithm we use to separate
the execution of single-partition and cross-partition
transactions. We first describe the two phases and how the
system transitions between them. 
We next give a brief proof to show that \codename produces serializable results.
In the end, we discuss how \codename achieves fault tolerance and recovers when failures occur.

\subsection{Partitioned Phase Execution} \label{ssec:partitioned_phase_execution}

Each node serves as primary for a subset of the records in the partitioned phase, as shown on the top of 
Figure~\ref{fig:phase_switching}. During this phase, we restrict the system to run transactions that only read from 
and write into a single partition. Cross-partition transactions
are deferred for later execution in the single-master phase.

In the partitioned phase, each partition is touched only by a single worker thread.
A transaction keeps a read set and a write set in its local copy, in case a transaction is aborted 
by the application explicitly. For example, an invalid item ID may be generated in TPC-C and a transaction 
with an invalid item ID is supposed to abort during execution. 
At commit time, it's not necessary to lock any record 
in the write set and do read validation, since there are no concurrent accesses to a partition in the partitioned phase.
The system still generates a TID for each transaction and uses the TID to tag the updated records.
In addition, writes of committed transactions are replicated to other backup nodes. 

\subsection{Single-Master Phase Execution} \label{ssec:single_master_phase_execution}

Any transaction can run in the single-master phase. 
Threads on the designated master node can access any record in any partition, since it has become the primary for all records.
For example, node 1 is the master node on the bottom of Figure~\ref{fig:phase_switching}.
We use multiple threads to run transactions using a variant of Silo's OCC protocol~\cite{TuZKLM13} in the single-master phase. 

A transaction reads data and the associated TIDs and keeps them in its read set for later read validation. 
During transaction execution, a write set is computed and kept in a local copy. 
At commit time, each record in the write set is locked
in a global order (e.g, the addresses of records) to prevent deadlocks. 
The transaction next generates a TID based on its read set, write set and the current epoch number. 
The transaction will abort during read validation if any record in the read set 
is modified (by comparing TIDs in the read set) or locked.  
Finally, records are updated, tagged with the new TID, and unlocked.
After the transaction commits, the system replicates its write set to other backup nodes. 

Note that fault tolerance must satisfy a transitive property~\cite{TuZKLM13}. 
The result of a transaction can only be released to clients after its writes and the writes of all transactions serialized before it are replicated. 
In \codename, the system treats each epoch as a commit unit through a replication fence. 
By doing this, it is guaranteed that  the serial order of transactions is always consistent with the epoch boundaries.  
Therefore, the system does not release the results of  transactions to clients until the next phase switch (and epoch boundary) occurs. 

\subsection{Phase Transitions} \label{ssec:phase_transitions}

We now describe how \codename transitions between the two phases, which alternate after the system  starts. 
The system starts in the partitioned phase, deferring 
cross-partition transactions for later execution.

For ease of presentation, we assume that all cross-partition transaction requests go to the designated master node 
(selected from among the $f$ nodes with a full copy of the database).
Similarly, each single-partition transaction request only goes to the participant node that has the mastership of the partition.
In the single-master phase, the master node becomes the primary for all records. 
In the partitioned phase, the master node acts like other participant nodes to run single-partition transactions.
This could be implemented via router nodes that are aware of the partitioning of the database.
If some transaction accesses multiple partitions on a non-master node, the system would re-route
the request to the master node for later execution.

In the partitioned phase, a thread on each partition 
fetches requests from clients and runs these transactions as discussed in Section~\ref{ssec:partitioned_phase_execution}. 
When the execution time in the partitioned phase exceeds a given threshold $\tau_p$, 
\codename switches all nodes into the single-master phase, as shown in Figure~\ref{alg:phase_switch_code}.
The phase switching algorithm is coordinated by a stand-alone {\it coordinator} outside of \codename instances. 
It can be deployed on any node of \codename or on a different node for better availability. 
 
Before the phase switching occurs, the coordinator in \codename stops all worker
threads.
During a replication fence, all participant nodes synchronize statistics about 
the number of committed transactions with one
another. From these statistics each node learns how many outstanding writes it
is waiting to see; nodes then wait until they have received and applied all
writes from the replication stream to their local database. 
Finally, the coordinator switches the system to the other phase.

In the single-master phase, 
worker threads on the master node pull requests from clients and run transactions 
as discussed in Section~\ref{ssec:single_master_phase_execution}. 
Meanwhile, the master node sends writes of committed transactions to replicas and 
all the other participant nodes stand by for replication. 
To further improve the utilization of servers, 
read-only transactions can run under read committed isolation level on non-master nodes at the client's discretion. 
Once the execution time in the single-master phase exceeds a given threshold $\tau_s$, the system switches back to the partitioned phase using another replication fence.

The parameters $\tau_p$ and $\tau_s$ are set dynamically according to
the system's throughput $t_p$ in the partitioned phase, the system's throughput $t_s$ in the single-master phase, 
the percentage $\mathcal{P}$ of cross-partition transactions in the workload, and  the iteration time $e$.
\begin{eqnarray}
   \tau_p + \tau_s & = & e\\
   \frac{\tau_s t_s} {\tau_p t_p + \tau_s  t_s}    & = &   \mathcal{P}
\end{eqnarray}
Note that $t_p$, $t_s$ and $\mathcal{P}$ are monitored and collected by the system in real time, and $e$ is supplied by the user.
Thus, these equations can be used to solve for $\tau_p$ and $\tau_s$, as these are the only two unknowns.
When there are no cross-partition transactions (i.e., $\mathcal{P} = 0$ and $t_s$ is not well-defined), the system sets $\tau_p$ to $e$ and $\tau_s$ to 0.

Intuitively, the system spends less time on synchronization with a longer iteration time $e$. 
In our experiments, we set the default iteration time to 10~ms; 
this provides good throughput while keeping latency at a typical level for high throughput transaction processing systems (e.g., Silo~\cite{TuZKLM13} uses 40~ms as a default).

Note that this deferral-based approach is symmetric so that single-partition transactions have the same expected mean latency as cross-partition transactions regardless of the iteration time (i.e., $\tau_p + \tau_s$),
assuming all transactions arrive at a uniform rate. 
For a transaction, the latency depends on when the phase in which it is going to run ends.
The mean latency is expected to be $(\tau_p + \tau_s) / 2$.

\subsection{Serializability} \label{ssec:serializability}

We now give a brief argument that transactions executed in \codename are serializable. 
A transaction only executes in a single phase, i.e., it runs in either the partitioned phase 
or the single-master phase. A replication fence between the partitioned phase and the single-master phase ensures that all writes 
from the replication stream have been applied to the database before switching to the next phase. 

In the partitioned phase, there is only one thread running transactions serially on each partition. Each executed transaction only touches one partition, which makes 
transactions clearly serializable. In the single-master phase, \codename implements a variant of Silo's OCC protocol~\cite{TuZKLM13} to ensure that
concurrent transactions are serializable. With the Thomas write rule, the secondary records are  correctly 
synchronized, even though the log entries from the replication stream may be applied in an arbitrary order.

\subsection{Fault Tolerance}  \label{ssec:fault_tolerance}

\begin{figure*}[!t]
    \begin{minipage}[!t]{0.48\linewidth}
    \centering
    \includegraphics[width=0.99\columnwidth]{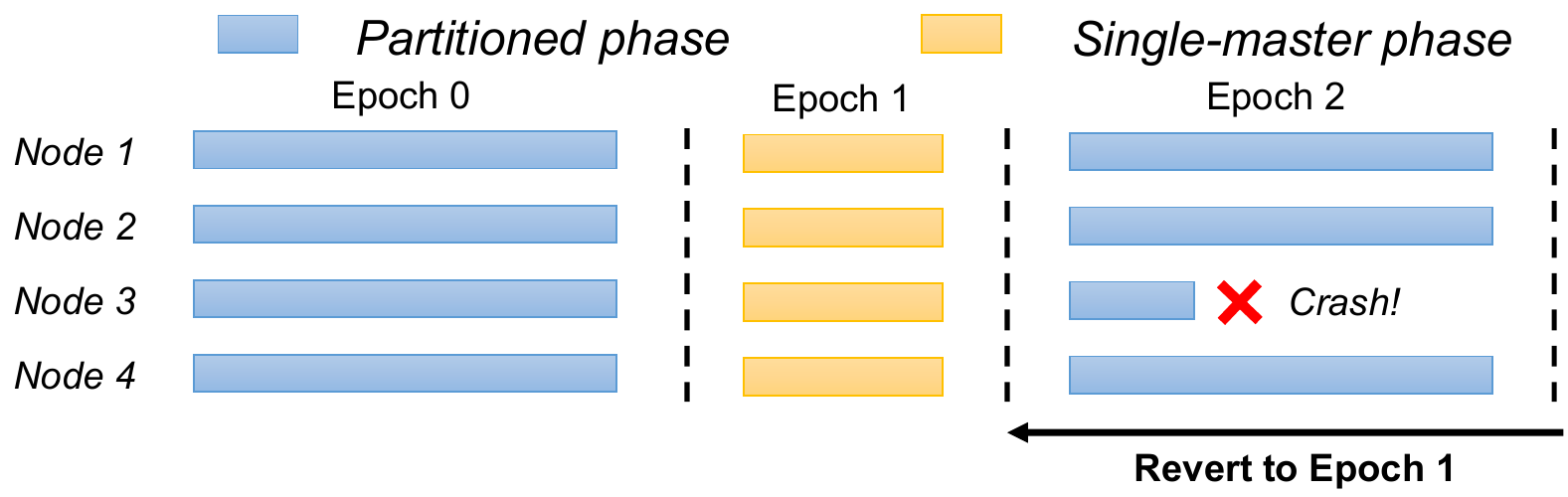}
    \vspace{-4mm}
    \caption{Failure detection in replication fence} \label{fig:failure_detection}
    \vspace{-4mm}
    \end{minipage}
    \hspace{0.02\linewidth}
    \begin{minipage}[!t]{0.48\linewidth}
    \centering
    \includegraphics[width=0.99\columnwidth]{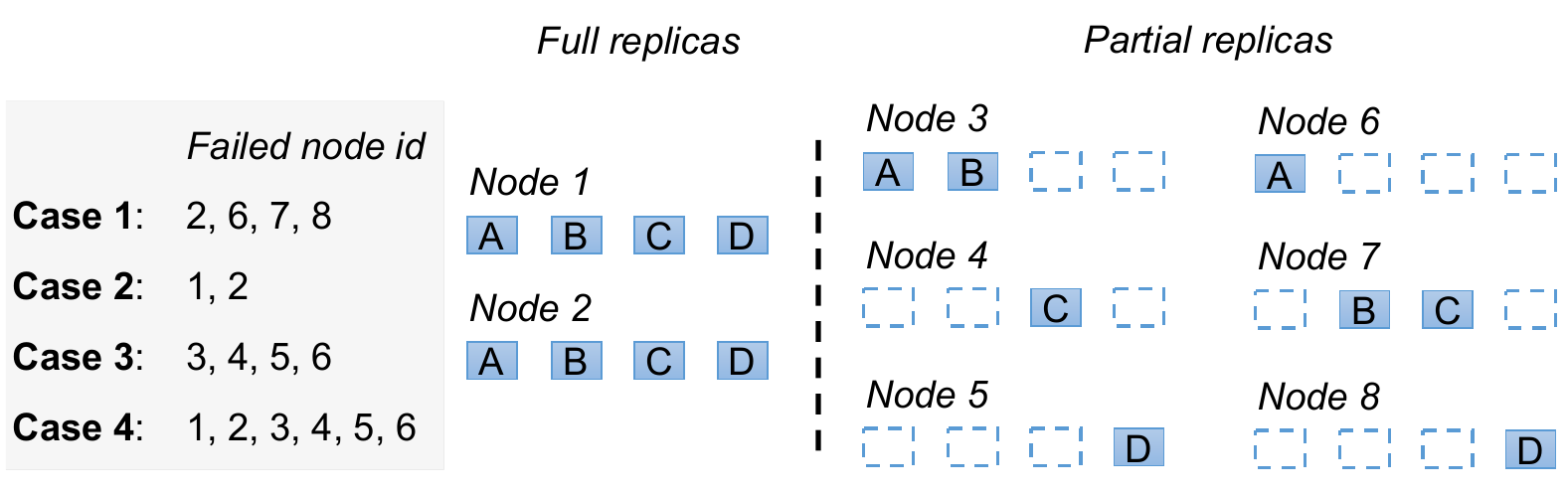}
    \vspace{-4mm}
    \caption{Illustrating different failure scenarios} \label{fig:fault_tolerance}
    \vspace{-4mm}    
    \end{minipage}
\end{figure*}

In-memory replication provides high availability to \codename, since transactions can run on other active nodes even though some nodes failed. 
To prevent data loss and achieve durability, the system must log writes of committed transactions to disk.
Otherwise, data will be lost when all replicas fail (e.g., due to power outage).

In this section, we first describe how \codename achieves durability via disk logging and checkpointing and then introduce
how failures are detected. Finally, we discuss how \codename recovers from failures under different scenarios. 

\subsubsection{Disk logging} \label{ssec:disk_logging}

In \codename, each worker thread has a local recovery log. 
The writes of committed transactions along with some metadata are buffered in memory and periodically flushed to the log.
Specifically, a log entry contains a single write to a record in the database, which has the following information: (1) key, (2) value, and (3) TID. 
The TID is from the transaction that last updated the record, 
and has an embedded epoch number as well. 
The worker thread periodically flushes buffered logs to disk;
\codename also flushes all buffers to disk in the replication fence. 

To bound the recovery time, a dedicated checkpointing thread can be used to periodically checkpoint the database to disk as well. 
The checkpointing thread scans the database and logs each record along with the TID to disk. 
A checkpoint also records the epoch number $e_c$ when it starts. 
Once a checkpoint finishes, all logs earlier than epoch $e_c$ can be safely deleted. 
Note that a checkpoint does not need to be a consistent snapshot of the database, as in SiloR~\cite{ZhengTKL14}, allowing the system to not freeze during checkpointing. 
On recovery, \codename uses the logs since the checkpoint (i.e., $e_c$) to correct the inconsistent snapshot with the Thomas write rule. 

\subsubsection{Failure detection} \label{ssec:dailure_detection}

Before introducing how \codename detects failures, we first give some definitions and assumptions on failures. 
In this paper, we assume fail-stop failures. 
A {\it healthy node} is a node that can connect to the coordinator, accept a client's requests, run transactions and replicate writes to other nodes. 
A {\it failed node} is one on which the process of a \codename instance has crashed~\cite{DeCandiaHJKLPSVV07} or which cannot communicate over the network.

The coordinator detects failures during the replication fence. 
If some node does not respond to the coordinator, it is considered to be a failed node. 
The list of failed nodes is broadcast to all healthy participant nodes in \codename.
In this way, healthy nodes can safely ignore all replication messages from failed nodes that have lost network connectivity to the coordinator. 
Thus, the coordinator acts as a view service to solve the ``split brain'' problem, coordinating data movement across nodes on failures. 
To prevent the coordinator from being a single point of failure, it can be implemented as a replicated state machine with Paxos~\cite{Lamport01} or Raft~\cite{OngaroO14}.

Once a failure is detected by the coordinator, the system enters recovery mode and reverts the database to the last committed epoch, as shown in Figure~\ref{fig:failure_detection}.
To achieve this, the database maintains two versions of each record. 
One is the most recent version prior to the current phase and the other one is the latest version written in the current phase.
The system ignores all data versions written in the current phase, since they have not been committed by the database.

We next describe how \codename recovers from failures once the database has been reverted to a consistent snapshot. 

\subsubsection{Recovery} \label{ssec:recovery}

We use examples from Figure~\ref{fig:fault_tolerance} to discuss how \codename recovers from failures. 
In these examples, there are 2 nodes with full replicas and 6 nodes with partial replicas (i.e., $f = 2$ and $k = 6$). 
A cluster of 8 nodes could fail in $2^8-1=255$ different ways, which fall into the following four different scenarios.  
Here, a ``full replica'' is a replica on a single node, and a ``complete partial replica'' is a set of partial replicas that collectively store a copy of the entire database.

\begin{enumerate}[label={(\arabic*)},noitemsep,nolistsep]
	\item At least one full replica and one complete partial replica remain.
	\item No full replicas remain but at least one complete partial replica remains.
	\item No complete partial replicas remain but at least one full replica remains.
	\item No full replicas or complete partial replicas remain.
\end{enumerate}

We now describe how \codename recovers under each scenario. 

{\noindent \bf Case 1:} As shown in Figure~\ref{fig:fault_tolerance}, failures occur on nodes 2, 6, 7 and 8. 
The system can still run transactions with the phase-switching algorithm. 
When a failed node recovers, it copies data from remote nodes and applies them to its database. 
In parallel, it processes updates from the relevant currently healthy nodes using  the Thomas write rule.
Once all failed nodes finish recovery, the system goes back to the normal execution mode.

{\noindent \bf Case 2:} If no full replicas are available, the system falls back to a mode in which a distributed concurrency control algorithm is employed, as in distributed partitioning-based systems (e.g., \occ). 
The recovery process on failed nodes is the same as in Case 1. 

{\noindent \bf Case 3:} If no complete partial replicas are available, the system can still run transactions with the phase-switching algorithm. 
However, the mastership of records on lost partitions have to be reassigned to the nodes with full replicas.
If all nodes with partial replicas fail, the system runs transactions only on full replicas without the phase-switching algorithm.
The recovery process on failed nodes is the same as in Case 1.

{\noindent \bf Case 4:} The system stops processing transactions (i.e., loss of availability) when no complete replicas remain. 
Each crashed node loads the most recent checkpoint from disk and restores its  database state to the end of the last epoch by replaying the logs since the checkpoint with the Thomas write rule.
The system goes back to the normal execution mode once all nodes finish recovery. 

Note that \codename also supports recovery from nested failures. 
For example, an additional failure on node 3 could occur during the recovery of Case 1. 
The system simply reverts to the last committed epoch and begins recovery as described in Case 3.

\section{Replication: Value vs. Operation} \label{sec:replication}

In this section, we describe the details of our replication schemes, and how replication is done depending on
the execution phase.
As discussed earlier, \codename runs single-partition and cross-partition 
transactions in different phases. The system uses different replication schemes in these two phases:
in the single-master phase, because a partition 
can be updated by multiple threads, records need to be fully-replicated to all replicas to ensure
correct replication.  However, in the partitioned phase, where a partition is only updated 
by a single thread, the system can use a better replication strategy based on replicating
operations to improve performance. \codename provides APIs for users to manually program the operations, e.g., string concatenation.

\begin{figure}[!t]
    \centering
    \includegraphics[width=0.99\columnwidth]{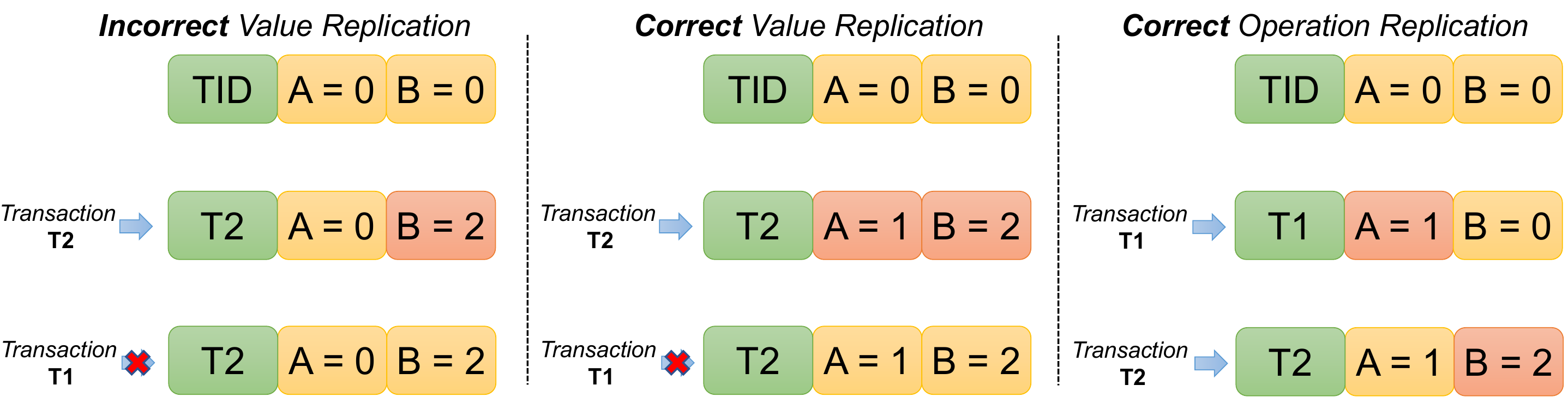}
        \vspace{-4mm}
    \caption{Illustrating different replication schemes; Red rectangle shows an updated field} \label{fig:replication_strategies}
        \vspace{-4mm}
\end{figure}

To illustrate this, consider two transactions being run by two threads: \texttt{T1: R1.A = R1.B + 1; R2.C = 0} 
and \texttt{T2: R1.B = R1.A + 1; R2.C = 1}. Suppose \texttt{R1} and \texttt{R2} are two records from
different partitions and we are running in the single-master phase.  
In this case, because the writes are done by different threads,
the order in which the writes arrive on replicas may be different from the order in which transactions commit on the primary.  
To ensure correctness, we employ the Thomas write rule: apply a write if the TID of the write is larger than the current TID of the record.
However, for this rule to work, each write must include the values of all fields in the record, not just
the updated fields.  To see this, consider the example in the left side of Figure~\ref{fig:replication_strategies} (only R1 is shown);
For record \texttt{R1}, if \texttt{T1} only replicates \texttt{A}, \texttt{T2} only replicates \texttt{B}, and \texttt{T2}'s updates 
are applied before \texttt{T1}'s, transaction \texttt{T1}'s update to field \texttt{A} is lost, 
since \texttt{T1} is less than \texttt{T2}. Thus, when a partition can be updated by multiple threads, 
all fields of a record have to be replicated as shown in the middle of Figure~\ref{fig:replication_strategies}. Note that fields that are always read-only do not need to be replicated.

Now, suppose \texttt{R1} and \texttt{R2} are from the same partition, and we run the same transactions in the partitioned phase, where transactions are run by only a single thread on each partition. If \texttt{T2} commits after \texttt{T1}, \texttt{T1} is guaranteed to 
be ahead of \texttt{T2} in the replication stream since they are executed by the same thread. 
For this reason, only the updated fields need to be replicated, i.e., \texttt{T1} can just send the new value for \texttt{A}, and \texttt{T2} can just send the new value for \texttt{B} as shown in the right side of Figure~\ref{fig:replication_strategies}. 
Furthermore, in this case, the system can also choose to replicate the \textit{operation} 
made to a field instead of the value of a field in a record. This can significantly reduce the amount of
data that must be sent.
For example, in the \texttt{Payment} transaction in TPC-C,  a string is concatenated 
to a field with a 500-character string in \texttt{Customer} table. 
With operation replication, the system only needs to replicate a short string and can re-compute the concatenated 
string on each replica, which is much less expensive than sending a 500-character string over network. 
This optimization can result in an order-of-magnitude reductions in replication cost.

In \codename, a \textit{hybrid replication} strategy is used, i.e., 
the master node uses value replication strategy in the single-master phase and 
all nodes use the operation replication strategy in the partitioned phase. 
The hybrid strategy achieves the best of of both worlds: 
(1) value replication enables out-of-order replication, not requiring a serial order which becomes a bottleneck in the single master phase (See Section~\ref{ssec:replication_and_fault_tolerance}), and 
(2) in the partitioned phase, operation replication reduces the communication cost compared to value replication, which always replicates the values of all fields in a record.

As discussed earlier, \codename logs the writes of committed transactions to disk for durability. 
The writes can come from either local transactions or remote transactions through replication messages. 
By default, \codename logs the whole record to disk for fast and parallel recovery. 
However, a replication message in operation replication only has operations rather than the value of a whole record. 
Consider the example in the right side of Figure~\ref{fig:replication_strategies}.
The replication messages only have \texttt{T1:}~\texttt{A = 1} and \texttt{T2:}~\texttt{B = 2}.
To solve this problem, when a worker thread processes a replication message that contains an operation, 
it first applies the operation to the database and then copies the value of the whole record to its logging buffer.
In other words, the replication messages are transformed into \texttt{T1:}~\texttt{A = 1; B = 0} and \texttt{T2:}~\texttt{A = 1; B = 2} before logging to disk.
By doing this, the logs can still be replayed in any order with the Thomas write rule during recovery.

\section{Discussion} \label{sec:discussion}

We now discuss the trade-offs that non-partitioned systems and partitioning-based systems achieve and use an analytical model to show how  \codename achieves the best of both worlds. 

\subsection{Non-partitioned Systems}

A typical approach to build a fault tolerant non-partitioned system is to adopt the primary/backup model.
A primary node runs transactions and replicates writes of committed transactions to one or more backup nodes.
If the primary node fails, one of the backup nodes can take over immediately without loss of availability. 

\begin{figure}
    \includegraphics[width=0.99\columnwidth]{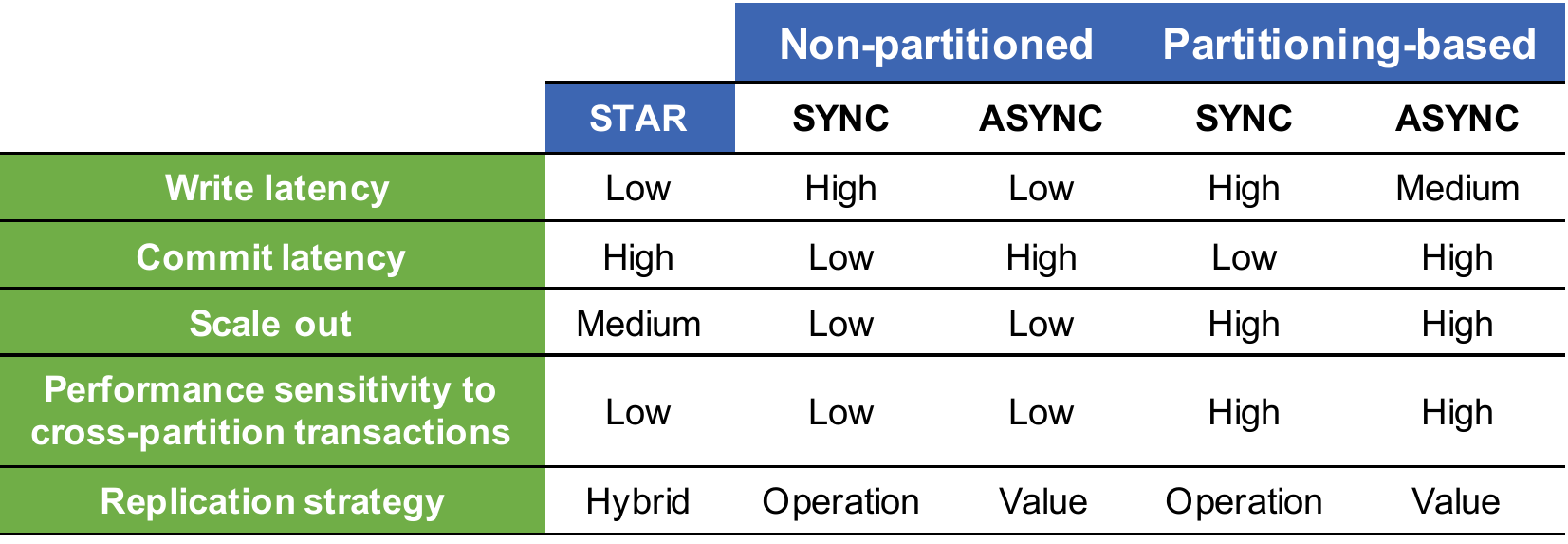}
    \vspace{-4mm}
    \caption{Overview of each approach; SYNC: synchronous replication; ASYNC: asynchronous replication + epoch-based group commit}\label{fig:overview}
    \vspace{-4mm}
\end{figure}

As we show in Figure~\ref{fig:overview}, the writes of committed transactions can be replicated from the primary node to backup nodes either synchronously or asynchronously. 
With synchronous replication, a transaction releases its write locks and commits as soon as the writes are replicated (low commit latency), however, round trip communication is needed even for single-partition transactions (high write latency). 
With asynchronous replication, it's not necessary to hold write locks on the primary node during replication, and the writes may be applied in any order on backup nodes with value replication and the Thomas write rule. 
To address the potential inconsistency issue when a fault occurs, an epoch-based group commit (high commit latency) must be used as well. The epoch-based group commit serves as a barrier that guarantees all writes are replicated when transactions commit. 
Asynchronous replication reduces the amount of time that a transaction holds write locks during replication (low write latency) but incurs high commit latency for all transactions.

The performance of non-partitioned systems has low sensitivity to cross-partition transactions in a workload, 
but they cannot easily scale out. 
The CPU resources on backup nodes are often under-utilized, using more hardware to provide a lower overall throughput. 

\subsection{Partitioning-based Systems}

In partitioning-based systems, the database is partitioned in a way such that each node owns one or more partitions. Each transaction has access to one or more partitions and commits with distributed concurrency control protocols (e.g., strict two-phase locking) and 2PC. This approach is a good fit for workloads that have a natural
partitioning as the database  can be treated as many disjoint sub-databases.
However, cross-partition transactions are frequent in  real-world scenarios.
For example, in the standard mix of \tpcc, 10\% of \texttt{NewOrder} and 15\% of \texttt{Payment} are cross-partition transactions.  

The same primary/backup model as in non-partitioned systems can be utilized to make partitioning-based systems fault tolerant. 
With synchronous replication, the write latency of partitioning-based systems is the same as non-par\-ti\-tioned systems. 
With asynchronous replication, the write latency depends on the number of partitions each transaction updates and on variance of communication delays.

If all transactions are single-partition transactions, par\-ti\-tioning-based systems are able to achieve linear scalability. 
However, even with a small fraction of cross-partition transactions, partitioning-based systems suffer from high round trip communication cost such as remote reads and distributed commit protocols.

\subsection{Achieving the Best of Both Worlds} \label{ssec:best_of_both_worlds}

\begin{figure}
    \includegraphics[width=0.99\columnwidth]{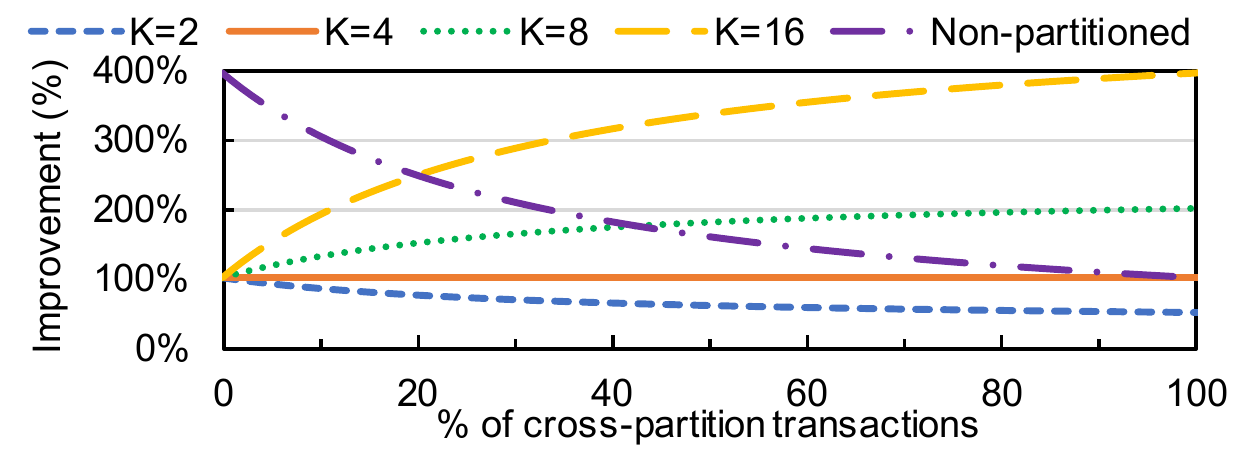}
    \vspace{-2mm}
    \caption{Illustrating effectiveness of \codename, vs. partitioning based systems for varying levels of $\mathcal{K}$, and against non-partitioned systems}\label{fig:model}
    \vspace{-4mm} 
\end{figure}

We now use an analytical model to show how  \codename achieves the best of both worlds. 
Suppose we have a workload with $n_s$ single-partition transactions and $n_c$ cross-partition transactions. 
We first analyze the time to run a workload with a partitioning-based approach on a cluster of $n$ nodes. 
If the average time of running a single-partition transaction and a cross-partition transaction in a partitioning-based system is $t_s$ and $t_c$ seconds respectively, we have
\begin{equation}
    T_{\text{Partitioning-based}}(n) = (n_s t_s + n_c t_c)/n     
\end{equation}
In contrast, the average time of running a cross-partition transaction is almost the same as running a single-partition transaction in a non-partitioned approach (e.g., in a primary-backup database), and therefore,
\begin{equation}
    T_{\text{Non-partitioned}}(n) = (n_s + n_c ) t_s
\end{equation}
In \codename, single-partition transactions are run on all participant nodes, and cross-partition transactions are run with a single master node. If the replication and phase transitions are not bottlenecks, we have
\begin{equation}
    T_{\text{\codename}}(n) = (n_s / n + n_c) t_s
\end{equation}
We let $\mathcal{K} = t_c / t_s $ and $\mathcal{P} = n_c / (n_c + n_s)$. 
Thus, $\mathcal{K}$ indicates how much more expensive a cross-partition transaction is than a single-partition transaction, 
and $\mathcal{P}$ indicates the percentage of cross-partition transactions in a workload. 
We now give the performance improvement that \codename achieves over the other two approaches,
\begin{eqnarray*}
I_{\text{Partitioning-based}}(n) & = & \frac{T_{\text{Partitioning-based}}(n)}{T_{\text{\codename}}(n)} =  \frac{\mathcal{K}\mathcal{P} - \mathcal{P} + 1}{n\mathcal{P} - \mathcal{P} + 1} \\
I_{\text{Non-partitioned}}(n) & = & \frac{T_{\text{Non-partitioned}}(n)}{T_{\text{\codename}}(n)} =  \frac{n}{n\mathcal{P} - \mathcal{P} + 1}
\end{eqnarray*}
Similarly, we have the scalability of asymmetric replication by showing 
the speedup that \codename achieves with $n$ nodes over a single node, 
\begin{displaymath}
     I(n) = \frac{T_{\text{\codename}}(1)}{T_{\text{\codename}}(n)} = \frac{n}{n \mathcal{P} - \mathcal{P} + 1}
\end{displaymath}
For different values of $\mathcal{K}$, we plot $I_{\text{Partitioning-based}}(4)$ and $I_{\text{Non-partitioned}}(4)$ in Figure~\ref{fig:model},
when varying the percentage of cross-partition transactions on a cluster of four nodes.
\codename outperforms non-partitioned systems as long as there are single-partition transactions in a workload. 
This is because all single-partition transactions are run on all participant nodes, 
which makes the system utilize more CPU resources from multiple nodes.
To outperform partitioning-based systems, the average time of running a cross-partition transaction must exceed $n$ times of the average time to run a single-partition transaction (i.e., $\mathcal{K} > n$). 

\section{Evaluation} \label{sec:evaluation}

In this section, we evaluate the performance of \codename focusing on the following key questions:

\begin{itemize}[noitemsep,nolistsep]
	\item How does \codename perform compared to non-partitioned systems and partitioning-based systems?
	\item How does \codename perform compared to deterministic databases?
	\item How does the phase switching algorithm affect the throughput of \codename and what's the overhead?
	\item How effective is \codename's replication strategy?
	\item How does \codename scale?
\end{itemize}

\subsection{Experimental Setup} \label{ssec:experimental_setup}

We ran our experiments on a cluster of four \texttt{m5.4xlarge} nodes running on Amazon EC2~\cite{ec2}.
Each node has 16 2.50 GHz virtual CPUs and 64 GB of DRAM running 64-bit Ubuntu 18.04.
\texttt{iperf} shows that the network between each node delivers about 4.8 Gbits/s throughput.
We implemented \codename and other distributed concurrency control algorithms in C++. 
The system is compiled using GCC 7.3.0 with \texttt{-O2} option enabled.

In our experiments, we run 12 worker threads on each node, yielding a total of 48 worker threads.
Each node also has 2 threads for network communication.
We made the number of partitions equal to the total number of worker threads.
All results are the average of three runs.
We ran transactions at the serializability isolation level.

\subsubsection{Workloads} \label{sssec:workloads}

To study the performance of \codename, we ran a number of experiments using two popular benchmarks:

{\bf YCSB:} The Yahoo! Cloud Serving Benchmark (YCSB) is a simple transactional workload designed to facilitate performance
comparisons of database and key-value systems~\cite{CooperSTRS10}. It has a single table with 10 columns.
The primary key of each record is a 64-bit integer and each column consists of 10
random bytes. 
A transaction accesses 10 records and each access follows a uniform distribution. 
We set the number of records to 200K per partition, and we run a workload mix of 90/10, 
i.e., each transaction has 9 read operations and 1 read/write operation. By default, we run this workload with 10\% cross-partition transactions that access to multiple partitions. 

{\bf TPC-C:} The TPC-C benchmark~\cite{tpcc}
is the gold standard for evaluating OLTP databases. It models a warehouse order processing system,
which simulates the activities found in complex OLTP applications. It has nine tables and
we partition all the tables by \texttt{Warehouse ID}.
We support two transactions in TPC-C, namely, (1) \texttt{NewOrder} and (2) \texttt{Payment}.
88\% of the standard TPC-C mix consists of these two transactions.
The other three transactions require range scans, which are currently not supported in our system.
By default, we ran this workload with the standard mix, 
in which a \texttt{NewOrder} transaction is followed by a \texttt{Payment} transaction. 
By default, 10\% of \text{NewOrder} and 15\% of \text{Payment} transactions are cross-partition transactions 
that access multiple warehouses. 

In YCSB, each partition adds about 25~MB to the database. In TPC-C, each partition contains one warehouse and adds about 100~MB to the database. 

To measure the maximum throughput that each approach can achieve, every worker thread generates and runs a transaction to completion one after another in our experiments.

\subsubsection{Concurrency control algorithms} \label{sssec:concurrency_control_algorithms}

To avoid an apples-to-oranges comparison, we implemented each of the following concurrency control algorithms in C++ in our framework.

{\bf \codename:} This is our algorithm as discussed in Section~\ref{sec:architecture}. 
We set the iteration time of a phase switch to 10~ms.
To have a fair comparison to other algorithms, disk logging, checkpointing, and the hybrid replication optimization are disabled unless otherwise stated.
    
{\bf \pb:} This is a variant of Silo's OCC protocol~\cite{TuZKLM13} adapted for a primary/backup setting. The primary node runs all transactions and replicates the writes to the backup node. Only two nodes are used in this setting.

{\bf \occ}: This is a distributed optimistic concurrency control protocol. 
A transaction reads from the database and maintains a local write set in the execution phase. 
The transaction first acquires all write locks and next validates all reads. 
Finally, it applies the writes to the database and releases the write locks.

{\bf \twopl:} This is a distributed strict two-phase locking protocol. 
A transaction acquires read and write locks during execution.
The transaction next executes to compute the value of each write. 
Finally, it applies the writes to the database and releases all acquired locks.

In our experiments, \pb is a non-partitioned system, and \occ and \twopl are considered as partitioning-based systems. We use \texttt{NO\_WAIT} policy to avoid deadlocks in partitioning-based systems, i.e., a transaction aborts if it fails to acquire some lock. This deadlock prevention strategy was shown to be the most scalable protocol~\cite{HardingAPS17}. We do not report the results on PB.~S2PL, since it always performs worse than \pb~\cite{YuPSD16}. Also note that we added an implementation of Calvin, described in Section~\ref{ssec:calvin}.

\subsubsection{Partitioning and replication configuration} \label{sssec:partitioning_and_replication_configuration}

In our experiment, we set the number of replicas of each partition to 2.
Each partition is assigned to a node by a hash function.
The primary partition and secondary partition are always hashed to two different nodes. 
In \codename, we have 1 node with full replica and\ 3 nodes with partial replica, i.e., $f = 1$ and $k = 3$. Each node masters 
a different portion of the database, as shown in Figure~\ref{fig:architecture}.

We consider two variations of \pb, \occ, and \twopl: (1) asynchronous replication and epoch-based group commit, and (2) synchronous replication.
Note that \occ and \twopl must use two-phase commit when synchronous replication is used. 
In addition, synchronous replication requires that all transactions hold write locks during the round trip communication for replication.

\subsection{Performance Comparison} \label{ssec:performance_comparison}

We now compare \codename with a non-partitioned system and two partitioning-based systems using both YCSB and TPC-C workloads.

\begin{figure*}[!t]
    \centering
    \subfigure[YCSB]{\includegraphics[width=0.49\columnwidth]{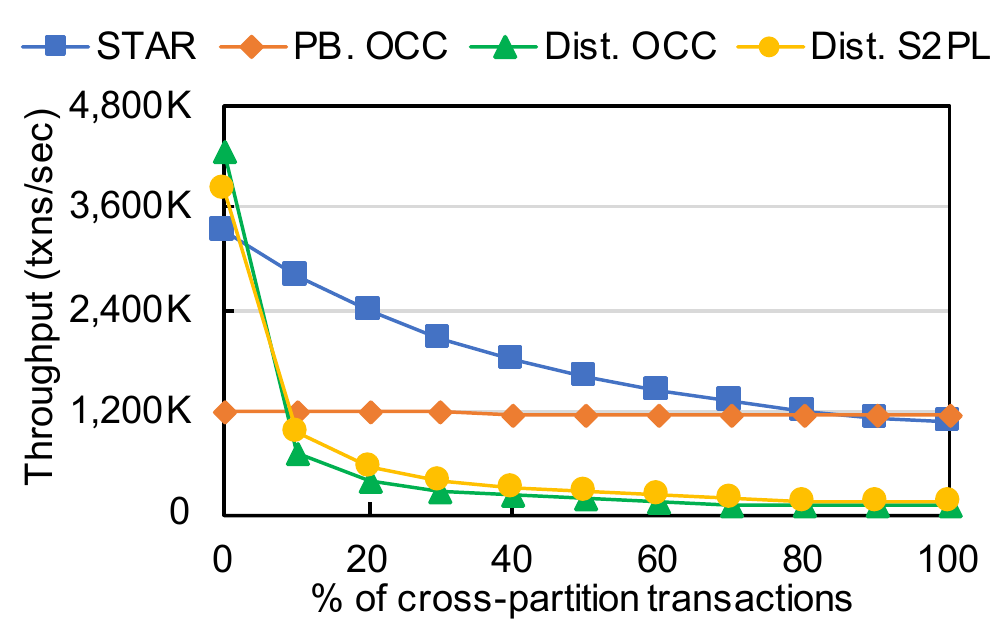}\label{fig:ycsb_async}}
    \subfigure[TPC-C]{\includegraphics[width=0.49\columnwidth]{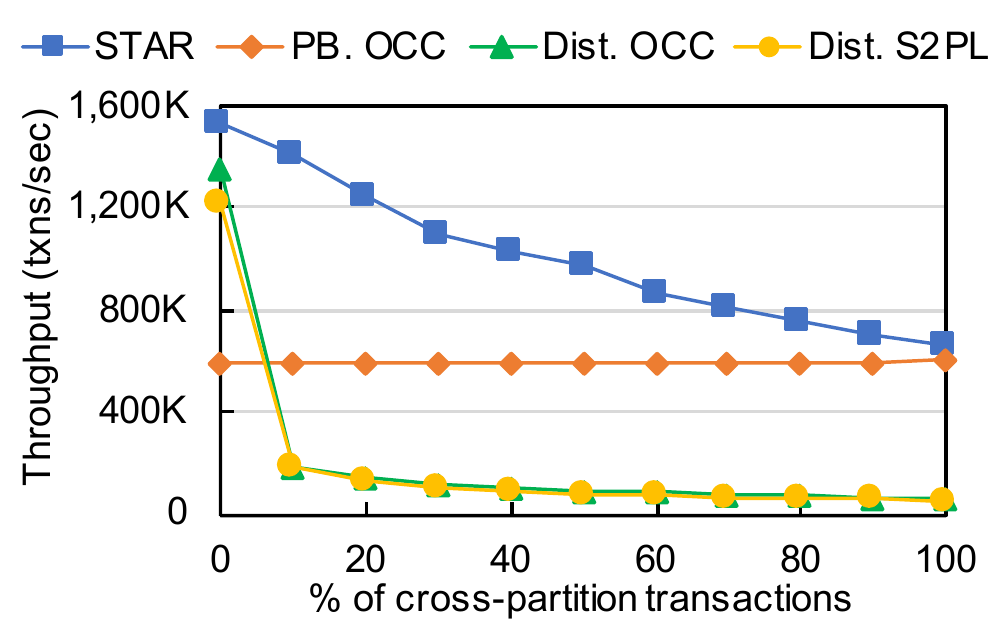}\label{fig:tpcc_async}}
    \subfigure[YCSB w/ Sync. Rep.]{\includegraphics[width=0.49\columnwidth]{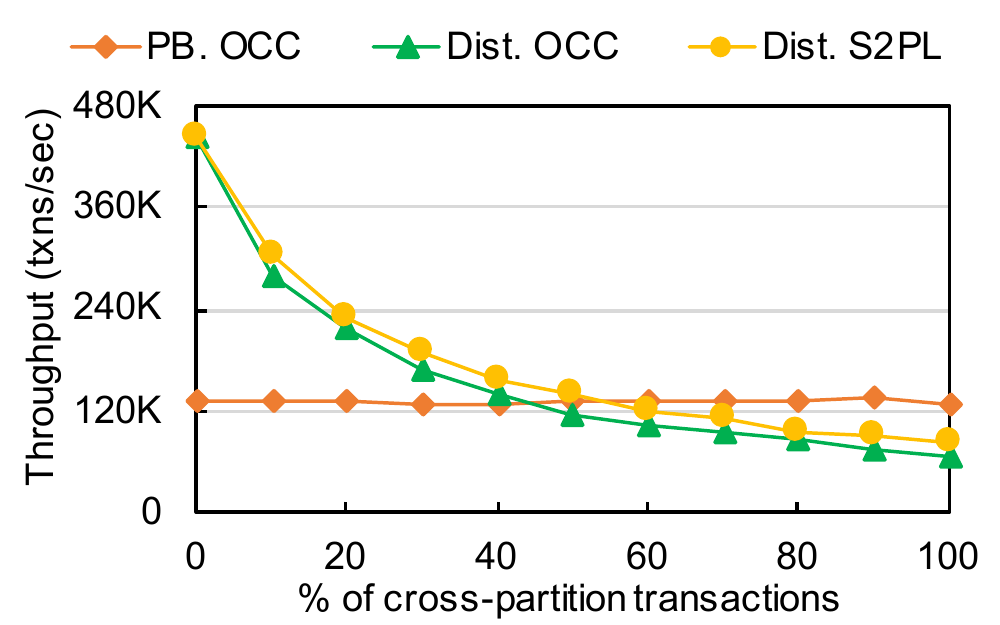}\label{fig:ycsb_sync}}
    \subfigure[TPC-C w/ Sync. Rep.]{\includegraphics[width=0.49\columnwidth]{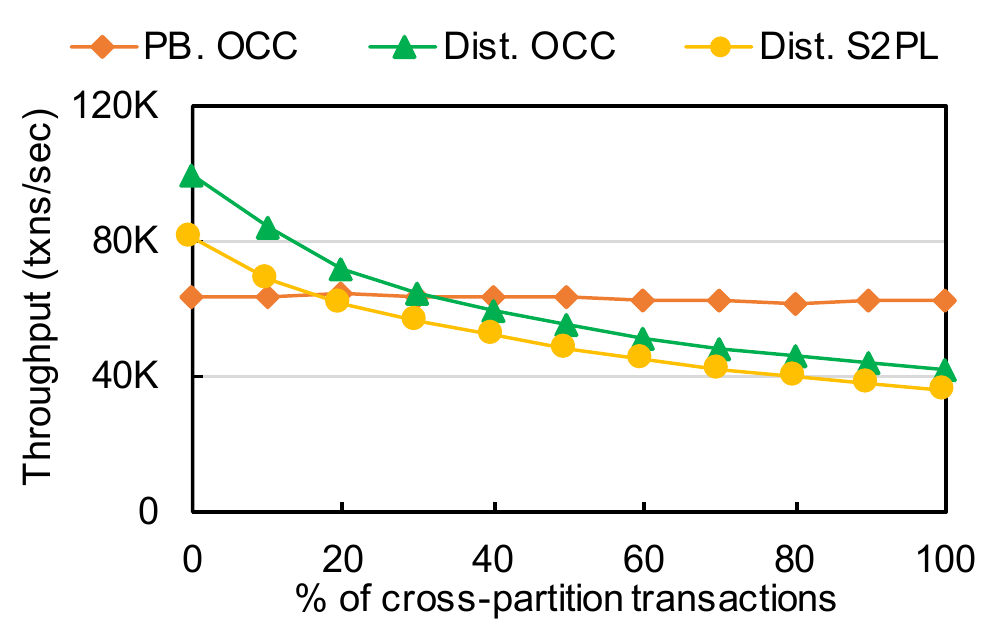}\label{fig:tpcc_sync}}
    \vspace{-4mm}
    \caption{Performance and latency comparison of each approach on YCSB and TPC-C} \label{fig:perf_and_latency}
\end{figure*}

\begin{figure*}[!t]
    \centering
    \includegraphics[width=1.99\columnwidth]{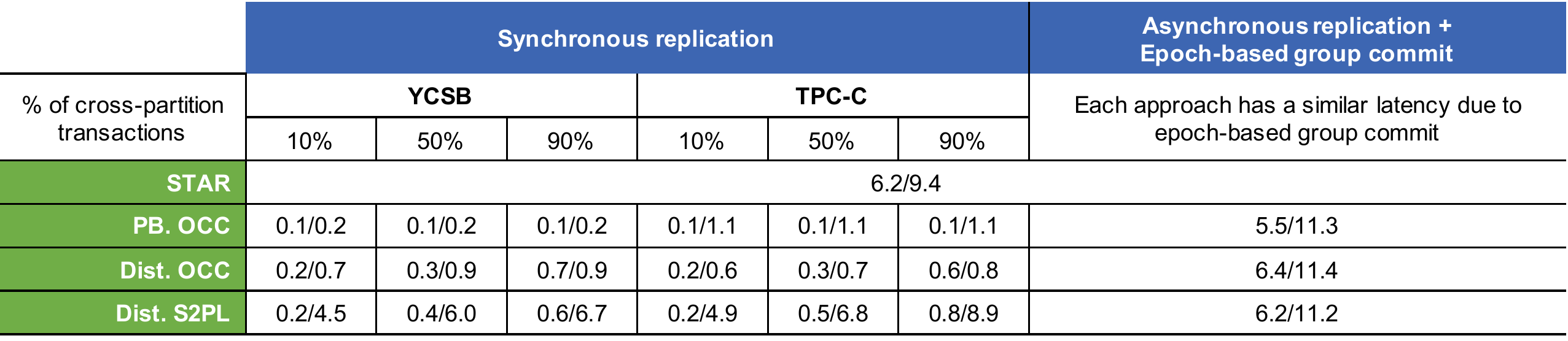}
    \vspace{-2mm}
    \caption{Latency (ms) of each approach - 50th percentile/99th percentile} \label{fig:latency}
    \vspace{-4mm}
\end{figure*}

\subsubsection{Results of asynchronous replication and epoch-based group commit} 
\label{sssec:group_commit}
 
We ran both YCSB and TPC-C with a varying percentage of cross-partition transactions and report the results in Figure~\ref{fig:ycsb_async} and~\ref{fig:tpcc_async}.
When there are no cross-partition transactions, \codename has similar throughput compared with \occ and \twopl on both workloads. This is because the workload is embarrassingly parallel. 
Transactions do not need to hold locks for a round trip communication with asynchronous replication and epoch-based group commit. 
As we increase the percentage of cross-partition transactions, the throughput of \pb stays almost the same, and the throughput of other approaches drops. 
When 10\% cross-partition transactions are present, 
\codename starts to outperform \occ and \twopl. 
For example, \codename has 2.9x higher throughput than \twopl on YCSB and 7.6x higher throughput than \occ on TPC-C.
As more cross-partition transactions are present, the throughput of \occ and \twopl is significantly lower than \codename (also lower than \pb), and the throughput of \codename approaches the throughput of \pb. This is because \codename behaves similarly to a non-partitioned system when all transactions are cross-partition transactions.

Overall, \codename running on 4 nodes achieves up to 3x higher throughput than a primary/backup system running on 2 nodes (e.g., \pb) and up to 10x higher throughput than systems employing distributed concurrency control algorithms (e.g., \occ and \twopl) on 4 nodes.
As a result, we believe that \codename is a good fit for workloads with both single-partition and cross-partition transactions. 
It can outperform both non-partitioned and partitioning-based systems, as we envisioned in Figure~\ref{fig:cartoon}.

\subsubsection{Results of synchronous replication} \label{sssec:synchronous_replication}

We next study the performance of \pb, \occ and \twopl with synchronous replication. 
We ran the same workload as in Section~\ref{sssec:group_commit} with a varying percentage of cross-partition transactions and report the results in Figure~\ref{fig:ycsb_sync} and~\ref{fig:tpcc_sync}. 
For clarity, the $y$-axis uses different scales (See Figure~\ref{fig:ycsb_async} and~\ref{fig:tpcc_async} for the results of \codename).
When there are no cross-partition transactions, the workload is embarrassingly parallel. 
However, \pb, \occ, and \twopl all have much lower throughput than \codename in this scenario.
This is because even single-partition transactions need to hold locks during the round trip communication due to synchronous replication.
As we increase the percentage of cross-partition transactions, 
we observe that the throughput of \pb stays almost the same, 
since the throughput of a non-partitioned system is not sensitive to the percentage of cross-partition transactions in a workload.
\occ and \twopl have lower throughput, since more transactions need to read from remote nodes during the execution phase. They also need multiple rounds of communication to validate and commit transactions (2PC).

Overall, the throughput of \pb, \occ, and \twopl is much lower than those with asynchronous replication and epoch-based group commit due to the overhead of network round trips for every transaction. 
\codename has much higher throughput than these approaches with synchronous replication --- at least 7x higher throughput on YCSB, 15x higher throughput on TPC-C.

\subsubsection{Latency of each approach} \label{sssec::latency}
We now study the latency of each approach and report the latency at the 50th percentile and the 99th percentile in Figure~\ref{fig:latency}, when the percentage of cross-partition transactions is 10\%, 50\%, and 90\%.
We first discuss the latency of each approach with synchronous replication.
We observe that \pb's latency at the 50th percentile and the 99th percentile is not sensitive to the percentage of cross-partition transactions. 
\occ and \twopl have higher latency at both the 50th percentile and the 99th percentile, as we increase the percentage of cross-partition transactions.
This is because there are more remote reads and the commit protocols they use need multiple round trip communication. 
In particular, the latency of \twopl at the 99th percentile is close to 10~ms on TPC-C.   
In \codename, the iteration time determines the latency of transactions.
Similarly, the latency of transactions in \occ and \twopl with asynchronous replication and epoch-based group commit depends on epoch size. 
For this reason, \codename has similar latency at the 50th percentile and the 99th percentile to other approaches with asynchronous replication.
In Figure~\ref{fig:latency}, we only report the results on YCSB with 10\% cross-partition transactions for systems with asynchronous replication. 
Results on other workloads are not reported, since they are all similar to one another.

An epoch-based group commit naturally increases latency. 
Systems with synchronous replication have lower latency, 
but Figure~\ref{fig:ycsb_sync} and~\ref{fig:tpcc_sync} show that they have much lower throughput as well, even if no cross-partition transactions are present. 
In addition, the latency at the 99th percentile in systems with synchronous replication is much longer under some scenarios (e.g., \twopl on TPC-C).
As prior work (e.g., Silo~\cite{TuZKLM13}) has argued, a few milliseconds more latency is not a problem for most transaction processing workloads, especially given throughput gains.

\subsection{Comparison with Deterministic Databases} \label{ssec:calvin}

\begin{figure*}[!t]
    \begin{minipage}[!t]{0.49\linewidth}
    \centering
    \subfigure[YCSB]{\includegraphics[width=0.49\columnwidth]{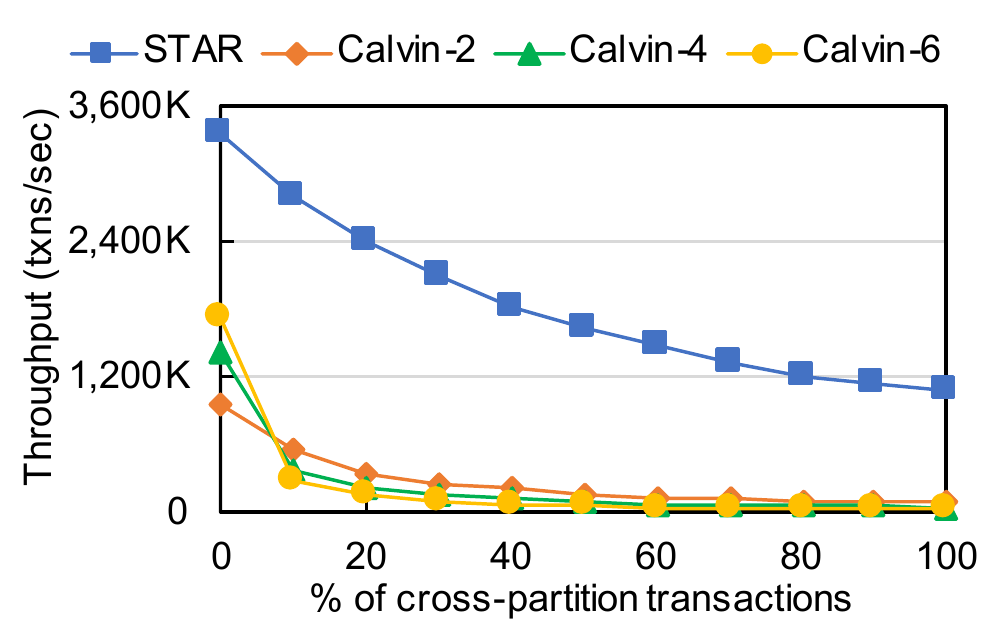}\label{fig:calvin_ycsb}}        
    \subfigure[TPC-C]{\includegraphics[width=0.49\columnwidth]{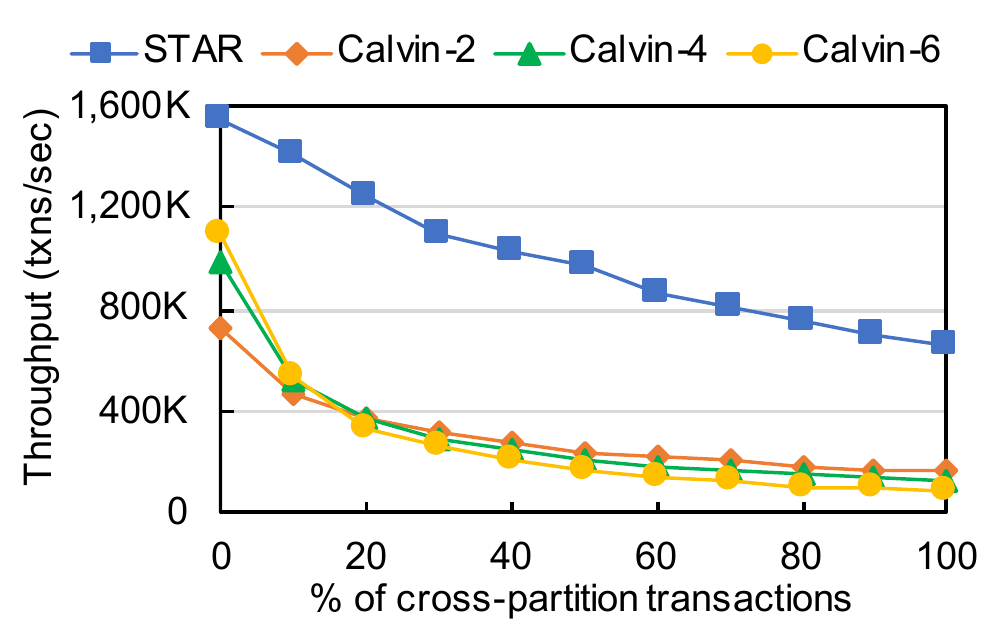}\label{fig:calvin_tpcc}}
    \vspace{-4mm}
    \caption{Comparison with deterministic databases} \label{fig:calvin}
    \vspace{-4mm}
    \end{minipage}
    \begin{minipage}[!t]{0.49\linewidth}
    \centering
    \subfigure[Iteration time (ms)]{\includegraphics[width=0.49\columnwidth]{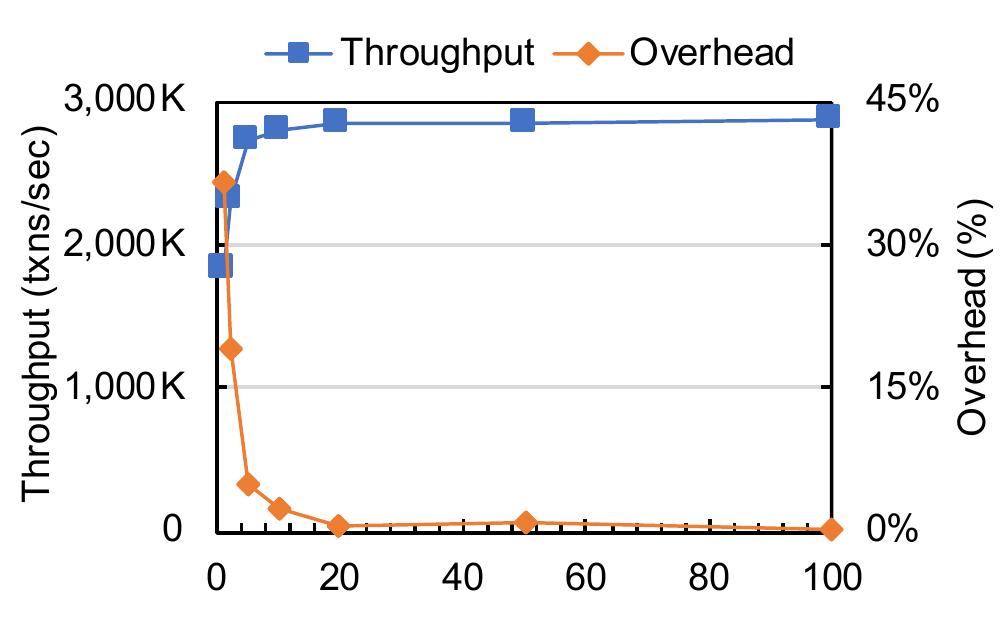}\label{fig:phase_transitions_epoch}}        
    \subfigure[\# of nodes]{\includegraphics[width=0.49\columnwidth]{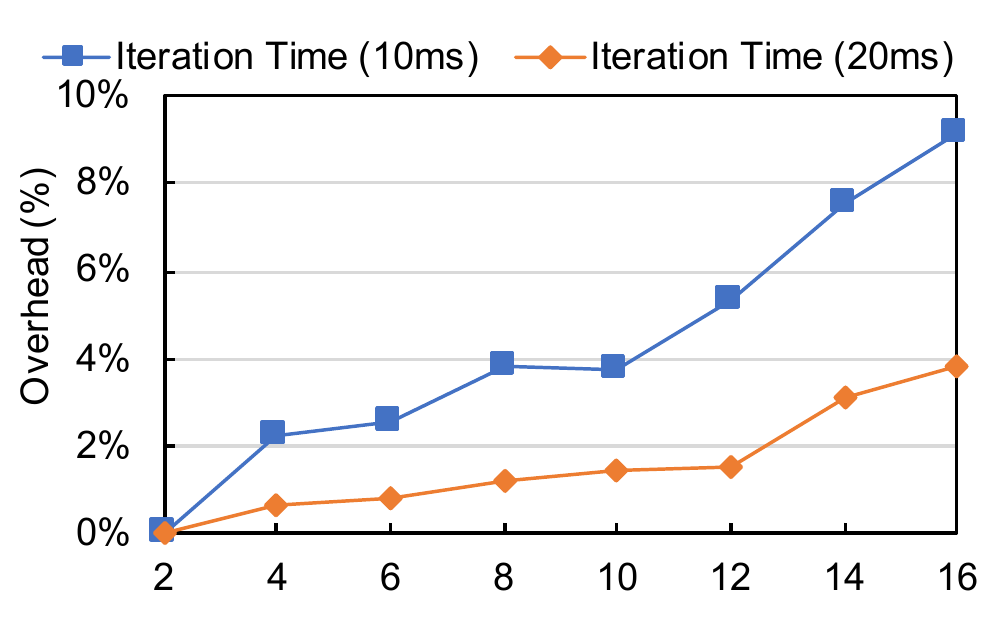}\label{fig:phase_transitions_scalability}}
    \vspace{-4mm}
    \caption{Overhead of phase transitions} \label{fig:phase_transitions}
    \vspace{-4mm}
    \end{minipage}
\end{figure*}

We next compare \codename with Calvin~\cite{ThomsonDWRSA12}, which is a deterministic concurrency control and replication algorithm.
In Calvin, a central sequencer determines the order for a batch of transactions before they start execution. 
The transactions are then sent to all replica groups of the database to execute deterministically.
In Calvin, a replica group is a set of nodes containing a replica of the database.
All replica groups will produce the same results for the same batch of transactions due to determinism. 
As a result, Calvin does not perform replication at the end of each transaction.  Instead, it replicates inputs at the beginning of the batch of transactions and deterministically executes the batch across replica groups.

We implemented the Calvin algorithm in C++ in our framework as well to have a fair comparison. 
The original design of Calvin uses a single-threaded lock manager to grant locks to multiple execution threads following the deterministic order. 
To better utilize more CPU resources, we implemented a multi-threaded lock manager, in which each thread 
is responsible for granting locks in a different portion of the database. 
The remaining CPU cores are used as worker threads to execute transactions.

Increasing the number of threads for the lock manager does not always improve performance. The reasons are twofold: (1) fewer threads are left for transaction execution (2) more communication is needed among worker threads for cross-partition transactions. 
In this experiment, we consider three configurations with different number of threads used in the lock manager in Calvin, namely (1) Calvin-2, (2) Calvin-4 and (3)  Calvin-6. 
We use Calvin-$x$ to denote the number of threads used for the lock manager, i.e, there are $12-x$ threads executing transactions. 
In all configurations, we study the performance of Calvin in one replica group on 4 nodes. 
The results of Calvin-1 and Calvin-3 are not reported, since they never deliver the best performance. 

We report the results on YCSB and TPC-C with a varying percentage of cross-partition transactions in Figure~\ref{fig:calvin_ycsb} and~\ref{fig:calvin_tpcc}. 
When there are no cross-partition transactions, Calvin-6 achieves the best performance, since more parallelism is exploited (i.e., 6 worker threads on each node, yielding a total of 24 worker threads). 
Calvin-2 and Calvin-4 have lower throughput as the worker threads are not saturated when fewer threads are used for the lock manager. 
In contrast, \codename uses 12 worker threads on each node, yielding a total of 48 worker threads and has 1.4-1.9x higher throughput than Calvin-6. 
When all transactions are cross-partition transactions, Calvin-2 has the best performance. 
This is because Calvin-4 and Calvin-6 needs more synchronization and communication.
Overall, \codename has 4-11x higher throughput than Calvin with various configurations.

\subsection{The Overhead of Phase Transitions} \label{ssec:the_overhead_of_phase_transitions}

We now study how the iteration time of a phase switch affects the overall throughput of \codename and
the overhead due to this phase switching algorithm with a YCSB workload.
Similar results were obtained on other workloads but are not reported due to space limitations.
We varied the iteration time of the phase switching algorithm from 1~ms to 100~ms and report the system's throughput and overhead in Figure~\ref{fig:phase_transitions_epoch}. 
The overhead is measured as the system's performance degradation compared to the one running with a 200~ms iteration time.
Increasing the iteration time decreases the overhead of the phase switching algorithm as expected,
since less time is spent during the synchronization. 
For example, when the iteration time is 1~ms, the overhead is as high as 43\% and system only achieves around half of its maximum throughput (i.e., the throughput achieved with 200~ms iteration time). 
As we increase the iteration time, the system's throughput goes up. 
The throughput levels off when the iteration time is larger than 10~ms. 
On a cluster of 4 nodes, the overhead is about 2\% with a 10 ms iteration time.

We also study the overhead of phase transitions with a varying number of nodes.
We ran the same YCSB workload and report the results of 10~ms and 20~ms iteration time in Figure~\ref{fig:phase_transitions_scalability}.
Note that we also scale the number of partitions in the database correspondingly. 
For example, on a cluster of 16 nodes, there are $16 \times 12 = 192$ partitions in the database.  
In general, the overhead of phase transitions is larger with more nodes on a cluster due to variance of communication delays.
In addition, a shorter iteration time makes the overhead smaller (20~ms vs. 10~ms). 

Overall, the overhead of phase transitions is less than 5\% with a 10~ms iteration time on a cluster of less than 10 nodes.
In all experiments in this paper, we set the iteration time to 10~ms. 
With this setting, the system can achieve more than 95\% of its maximum throughput 
and have good balance between throughput and latency.

\subsection{Replication and Fault Tolerance} \label{ssec:replication_and_fault_tolerance}

\begin{figure*}[!t]
    \begin{minipage}[!t]{0.49\linewidth}
    \centering
    \subfigure[Replication strategies]{\includegraphics[width=0.49\columnwidth]{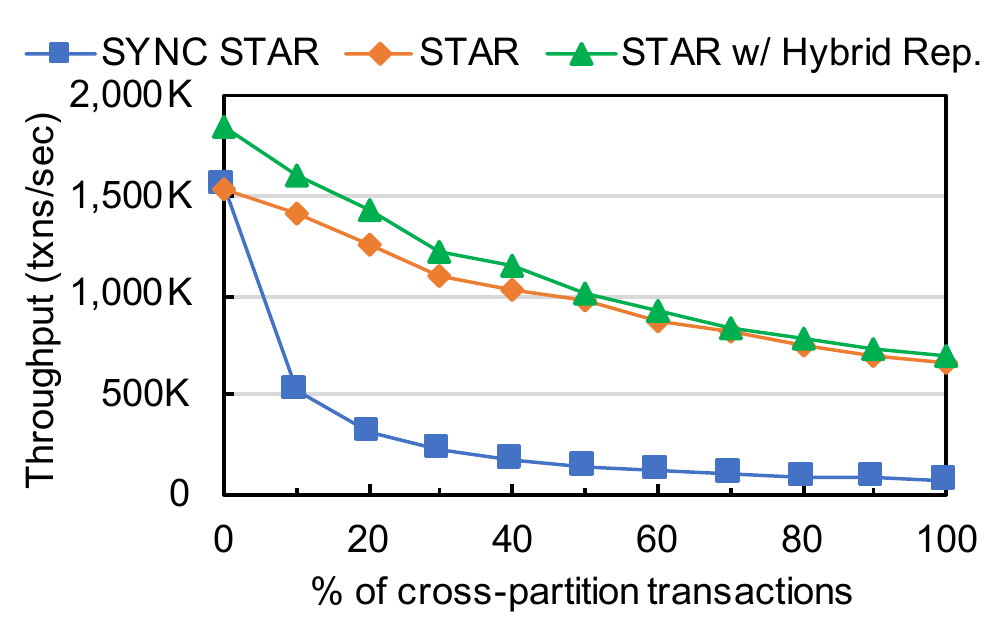}\label{fig:replication}}     
    \subfigure[Fault tolerance]{\includegraphics[width=0.49\columnwidth]{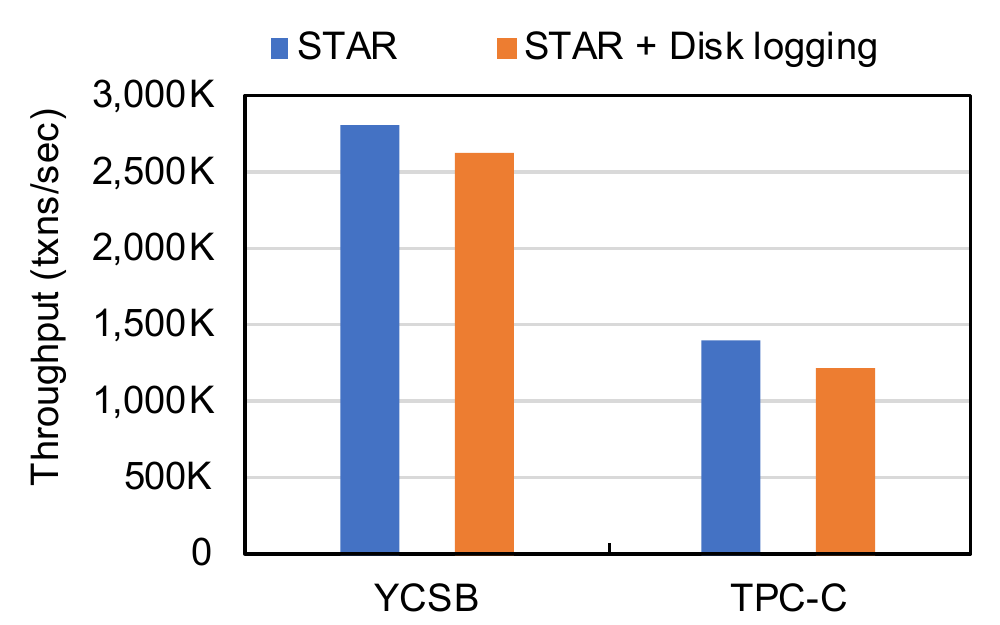}\label{fig:disk_logging}}
    \vspace{-4mm}
    \caption{Replication and fault tolerance experiment} \label{fig:replication_and_fault_tolerance}
    \vspace{-2mm}
    \end{minipage}
    \begin{minipage}[!t]{0.49\linewidth}
    \centering
    \subfigure[YCSB]{\includegraphics[width=0.49\columnwidth]{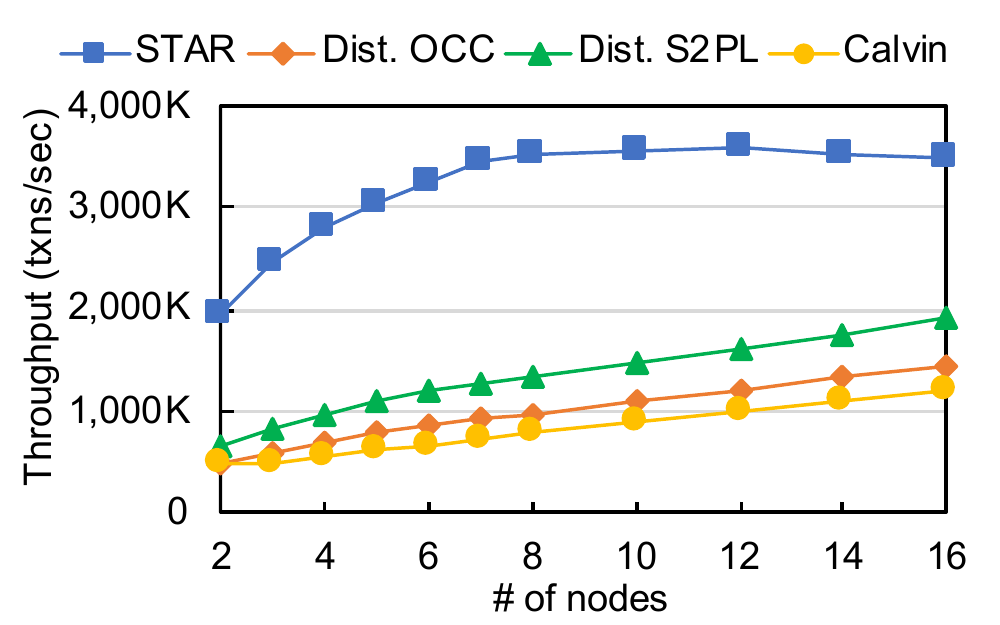}\label{fig:scalability_ycsb}}  
    \subfigure[TPC-C]{\includegraphics[width=0.49\columnwidth]{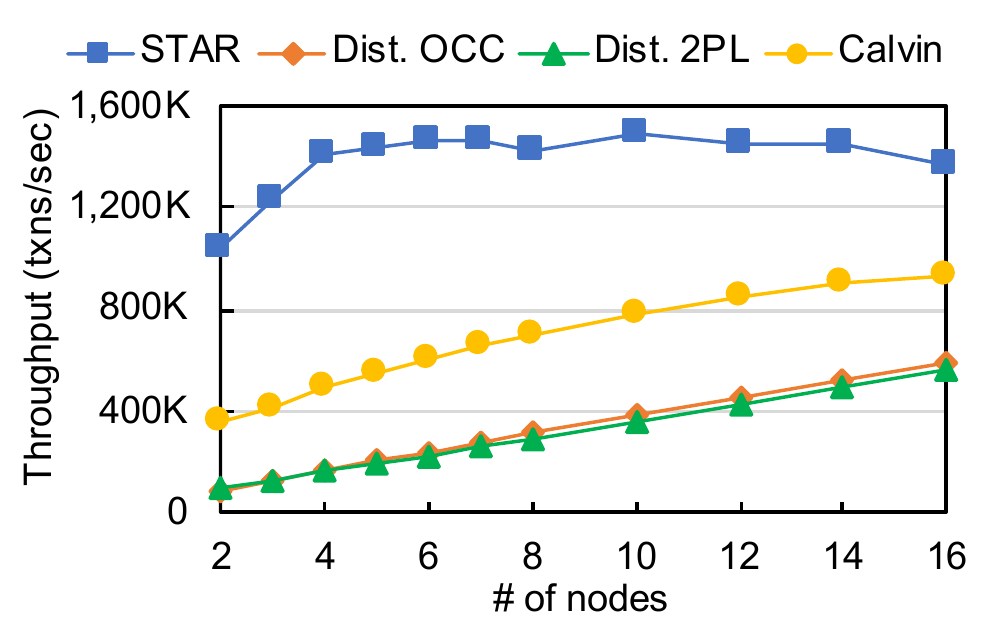}\label{fig:scalability_tpcc}}
    \vspace{-4mm}
    \caption{Scalability experiment} \label{fig:scalability}
    \vspace{-2mm}
    \end{minipage}
\end{figure*}

We now study the effectiveness of \codename's asynchronous replication in the single-master phase and the effectiveness of hybrid replication. 
We only report the results from TPC-C in this experiment, since a transaction in YCSB always updates the whole record.
In Figure~\ref{fig:replication}, \texttt{SYNC STAR} shows the performance of \codename that uses synchronous replication in the single-master phase. 
\texttt{STAR} indicates the one with asynchronous replication.
\texttt{STAR w/ Hybrid Rep.} further enables operation replication  
in the partitioned phase on top of \texttt{STAR}.
When there are more cross-partition transactions, 
\texttt{SYNC STAR} has much lower throughput than \texttt{STAR}.
This is because more network round trips are needed during replication in the single-master phase. 
The improvement of \texttt{STAR w/ Hybrid Rep.} is also less significant, since fewer transactions are run in the partitioned phase.

We next show the performance degradation of \codename when disk logging is enabled. 
We ran both YCSB and TPC-C workloads and report the results in Figure~\ref{fig:disk_logging}.  
In summary, the overhead of disk logging and checkpointing is 6\% in YCSB and 14\% in TPC-C. 
Note that non-partitioned and partitioning-based systems would experience similar overheads from disk logging.

\subsection{Scalability Experiment} \label{ssec:scalability}

In this experiment, we study the scalability of \codename on both YCSB and TPC-C.
We ran the experiment with a varying number of \texttt{m5.4xlarge} nodes and report the results in Figure~\ref{fig:scalability}.
Note that the database is scaled correspondingly as we did in Section~\ref{ssec:the_overhead_of_phase_transitions}.
On YCSB, \codename with 8 nodes achieves 1.8x higher throughput than \codename with 2 nodes. 
The performance of \codename stays stable beyond 8 nodes.
On TPC-C, \codename with 4 nodes achieves 1.4x higher throughput than \codename with 2 nodes.
The system stops scaling with more than 4 nodes. 
This is because the system saturates the network with 4 nodes (roughly 4.8 Gbits/sec).
In contrast, \occ, \twopl and Calvin start with lower performance but all have almost linear scalability. 

We believe it is possible for distributed partitioning-based systems to have competitive performance to \codename, although such systems will likely require more nodes to achieve comparable performance. 
Assuming linear scalability and the network not becoming a bottleneck (the ideal case for baselines), 
distributed partitioning-based systems maybe outperform \codename on YCSB and TPC-C with roughly 30-40 nodes.
\section{Related Work} \label{sec:related_work}

\codename builds on a number of pieces of related work for its design, including in-memory transaction processing, replication and durability.


{\it In-memory Transaction Processing.} Modern fast in-memory databases have long been an active area of 
research~\cite{Crooks0CHAA18, DingKG18, LarsonBDFPZ11, NeumannMK15, StonebrakerMAHHH07, TuZKLM13, YuPSD16}. 
In H-Store~\cite{StonebrakerMAHHH07}, transactions local to a single partition are executed by 
a single thread associated with the partition. This extreme form of partitioning makes single-partition 
transactions very fast but creates significant contention for cross-partition transactions, 
where whole-partition locks are held.
Silo~\cite{TuZKLM13} divides time into a series of short epochs and each thread generates its 
own timestamp by embedding the global epoch to avoid shared-memory writes, avoiding contention
for a global critical section.
Because of its high throughput and simple design, 
we adopted the Silo architecture for \codename, reimplementing it and 
adding our phase-switching protocol and replication. 
Doppel~\cite{NarulaCKM14} executes highly contentious transactions in a separate phase from other regular transactions such that special optimizations (i.e., commutativity) can be applied to improve scalability. 

F1~\cite{ShuteVSHWROLMECRSA13} is an OCC protocol built on top of Google's Spanner~\cite{CorbettDEFFFGGHHHKKLLMMNQRRSSTWW12}. 
MaaT~\cite{MahmoudANAA14} reduces transaction conflicts with dynamic timestamp ranges.
ROCOCO~\cite{MuCZLL14} tracks conflicting constituent pieces of transactions and re-orders them in a serializable order before execution. To reduce the conflicts of distributed transactions, \codename runs all cross-partition transactions on a single machine in the single-master phase.
Clay~\cite{SerafiniTEPAS16} improves data locality to reduce the number of distributed transactions in a distributed OLTP system by smartly partitioning and migrating data across the servers. 
Some previous work~\cite{LinC0OTW16, DasAA10, ChairunnandaDO14} proposed to move the master node of a tuple dynamically, in order to convert distributed transactions into local transactions. Unlike \codename, however, moving the mastership still requires network communication.
FaRM~\cite{DragojevicNCH14}, FaSST~\cite{KaliaKA16} and DrTM~\cite{WeiSCCC15} improve the performance of a distributed OLTP database by exploiting RDMA. 
\codename can use RDMA to further decrease the overhead of replication and the phase switching algorithm as well.

{\it Replicated Systems.} Replication is the way in which database
systems achieve high availability. 
Synchronous replication was popularized by systems like
Postgres-R~\cite{KemmeA00} and Galera Cluster~\cite{GaleraCluster}, which
showed how to make synchronous replication practical using group communication and
deferred propagation of writes.
Tashkent~\cite{ElniketyDP06} is a fully replicated database in which transactions run locally on a replica.
To keep replicas consistent, each replica does not communicate with each other but communicates to a certifier, which decides a global order for update transactions.
Calvin~\cite{ThomsonDWRSA12} replicates transactions requests 
among replica groups and assigns a global order to each transaction for 
deterministic execution~\cite{ThomsonA10}, allowing it to eliminate expensive distributed coordination. 
However, cross-node communication is still necessary during transaction execution 
because of remote reads. 
Mencius~\cite{MaoJM08} is a state machine replication method that improves Paxos to achieve high throughput under high client load and low latency under low client load  by partitioning sequence numbers, even under changing wide-area network environments. 
HRDB~\cite{VandiverBLM07} tolerates Byzantine faults among replicas by scheduling transactions with a commit barrier.
Ganymed~\cite{PlattnerA04, PlattnerAO08} runs update transactions on a single node and runs read-only transactions on a potentially unlimited number of replicas, allowing the system to scale read-intensive workloads. 
\codename is the first system that dynamically changes the mastership of records, to avoid distributed coordination.
Neither a global order nor group communication is necessary, even for cross-partition transactions, since we run these cross-partition transactions in parallel on a single node.


{\it Recoverable Systems.} 
H-Store~\cite{MalviyaWMS14} uses 
transaction-level logging. It periodically checkpoints a transactionally consistent snapshot to disk 
and logs all the parameters of stored procedures. H-Store executes transactions following a global order 
and replays all the transactions in the same order during recovery. SiloR~\cite{ZhengTKL14} uses a multi-threaded parallel value logging scheme that supports
parallel replay  in non-partitioned databases. In contrast, transaction-level 
logging requires that transactions be replayed in the same order. In \codename, different replication strategies, including both SiloR-like parallel value replication and H-Store-like operation replication are used
in different phases, significantly reducing bandwidth requirements.

\section{Conclusion} \label{sec:conclusion}

In this paper, we presented \codename, a new distributed in-memory 
database with asymmetric replication.
\codename employs a new \textit{phase-switching} 
scheme where single-partition transactions are run on multiple
machines in parallel, and cross-partition transactions are run on a single machine by
re-mastering records on the fly, allowing us to avoid cross-node communication 
and the use of distributed commit protocols 
like 2PC for distributed transactions.  
Our results on YCSB and TPC-C show
that \codename is able to dramatically exceed the
performance of systems that employ conventional concurrency control and replication algorithms by up to one order of magnitude.

\vspace{1mm}

{\noindent \bf Acknowledgments.} We thank the reviewers for their valuable comments. Yi Lu is supported by the Facebook PhD Fellowship.
\balance
\bibliographystyle{abbrv}
\bibliography{references}

\begin{thebibliography}{10}

\bibitem{tpcc}
{TPC Benchmark C}.
\newblock \url{http://www.tpc.org/tpcc/}, 2010.

\bibitem{ec2}
{Amazon EC2}.
\newblock \url{https://aws.amazon.com/ec2/}, 2019.

\bibitem{google-cloud}
{Google Cloud}.
\newblock \url{https://cloud.google.com/}, 2019.

\bibitem{BernsteinHG87}
P.~A. Bernstein, V.~Hadzilacos, and N.~Goodman.
\newblock {\em Concurrency Control and Recovery in Database Systems}.
\newblock Addison-Wesley, 1987.

\bibitem{ChairunnandaDO14}
P.~Chairunnanda, K.~Daudjee, and M.~T. {\"{O}}zsu.
\newblock {ConfluxDB}: Multi-master replication for partitioned snapshot
  isolation databases.
\newblock {\em {PVLDB}}, 7(11):947--958, 2014.

\bibitem{CooperSTRS10}
B.~F. Cooper, A.~Silberstein, E.~Tam, R.~Ramakrishnan, and R.~Sears.
\newblock Benchmarking cloud serving systems with {YCSB}.
\newblock In {\em SoCC}, pages 143--154, 2010.

\bibitem{CorbettDEFFFGGHHHKKLLMMNQRRSSTWW12}
J.~C. Corbett, J.~Dean, M.~Epstein, A.~Fikes, C.~Frost, J.~J. Furman,
  S.~Ghemawat, A.~Gubarev, C.~Heiser, P.~Hochschild, W.~C. Hsieh, S.~Kanthak,
  E.~Kogan, H.~Li, A.~Lloyd, S.~Melnik, D.~Mwaura, D.~Nagle, S.~Quinlan,
  R.~Rao, L.~Rolig, Y.~Saito, M.~Szymaniak, C.~Taylor, R.~Wang, and
  D.~Woodford.
\newblock Spanner: Google's globally-distributed database.
\newblock In {\em OSDI}, pages 261--264, 2012.

\bibitem{Crooks0CHAA18}
N.~Crooks, M.~Burke, E.~Cecchetti, S.~Harel, R.~Agarwal, and L.~Alvisi.
\newblock {Obladi}: Oblivious serializable transactions in the cloud.
\newblock In {\em OSDI}, pages 727--743, 2018.

\bibitem{DasAA10}
S.~Das, D.~Agrawal, and A.~{El Abbadi}.
\newblock {G-Store}: a scalable data store for transactional multi key access
  in the cloud.
\newblock In {\em SoCC}, pages 163--174, 2010.

\bibitem{DeBrabantPTSZ13}
J.~DeBrabant, A.~Pavlo, S.~Tu, M.~Stonebraker, and S.~B. Zdonik.
\newblock Anti-caching: {A} new approach to database management system
  architecture.
\newblock {\em {PVLDB}}, 6(14):1942--1953, 2013.

\bibitem{DeCandiaHJKLPSVV07}
G.~DeCandia, D.~Hastorun, M.~Jampani, G.~Kakulapati, A.~Lakshman, A.~Pilchin,
  S.~Sivasubramanian, P.~Vosshall, and W.~Vogels.
\newblock Dynamo: {Amazon}'s highly available key-value store.
\newblock In {\em SOSP}, pages 205--220, 2007.

\bibitem{DiaconuFILMSVZ13}
C.~Diaconu, C.~Freedman, E.~Ismert, P.~Larson, P.~Mittal, R.~Stonecipher,
  N.~Verma, and M.~Zwilling.
\newblock Hekaton: {SQL} server's memory-optimized {OLTP} engine.
\newblock In {\em {SIGMOD} Conference}, pages 1243--1254, 2013.

\bibitem{DingKG18}
B.~Ding, L.~Kot, and J.~Gehrke.
\newblock Improving optimistic concurrency control through transaction batching
  and operation reordering.
\newblock {\em {PVLDB}}, 12(2):169--182, 2018.

\bibitem{DragojevicNCH14}
A.~Dragojevic, D.~Narayanan, M.~Castro, and O.~Hodson.
\newblock Farm: Fast remote memory.
\newblock In {\em NSDI}, pages 401--414, 2014.

\bibitem{ElniketyDP06}
S.~Elnikety, S.~G. Dropsho, and F.~Pedone.
\newblock Tashkent: uniting durability with transaction ordering for
  high-performance scalable database replication.
\newblock In {\em EuroSys}, pages 117--130, 2006.

\bibitem{EswarranGLT76}
K.~P. Eswaran, J.~Gray, R.~A. Lorie, and I.~L. Traiger.
\newblock The notions of consistency and predicate locks in a database system.
\newblock {\em Commun. {ACM}}, 19(11):624--633, 1976.

\bibitem{FaleiroA15}
J.~M. Faleiro and D.~J. Abadi.
\newblock Rethinking serializable multiversion concurrency control.
\newblock {\em {PVLDB}}, 8(11):1190--1201, 2015.

\bibitem{GaleraCluster}
{Galera Cluster}.
\newblock \url{http://galeracluster.com/products/technology/}, 2019.

\bibitem{HardingAPS17}
R.~Harding, D.~V. Aken, A.~Pavlo, and M.~Stonebraker.
\newblock An evaluation of distributed concurrency control.
\newblock {\em {PVLDB}}, 10(5):553--564, 2017.

\bibitem{KaliaKA16}
A.~Kalia, M.~Kaminsky, and D.~G. Andersen.
\newblock Fasst: Fast, scalable and simple distributed transactions with
  two-sided {(RDMA)} datagram rpcs.
\newblock In {\em OSDI}, pages 185--201, 2016.

\bibitem{KemmeA00}
B.~Kemme and G.~Alonso.
\newblock Don't be lazy, be consistent: {Postgres-R}, {A} new way to implement
  database replication.
\newblock In {\em VLDB}, pages 134--143, 2000.

\bibitem{KimWJP16}
K.~Kim, T.~Wang, R.~Johnson, and I.~Pandis.
\newblock {ERMIA:} fast memory-optimized database system for heterogeneous
  workloads.
\newblock In {\em {SIGMOD} Conference}, pages 1675--1687, 2016.

\bibitem{KungR81}
H.~T. Kung and J.~T. Robinson.
\newblock On optimistic methods for concurrency control.
\newblock {\em {ACM} Trans. Database Syst.}, 6(2):213--226, 1981.

\bibitem{Lamport01}
L.~Lamport.
\newblock Paxos made simple.
\newblock {\em ACM SIGACT News}, 32(4):18--25, 2001.

\bibitem{LarsonBDFPZ11}
P.~Larson, S.~Blanas, C.~Diaconu, C.~Freedman, J.~M. Patel, and M.~Zwilling.
\newblock High-performance concurrency control mechanisms for main-memory
  databases.
\newblock {\em {PVLDB}}, 5(4):298--309, 2011.

\bibitem{LimKA17}
H.~Lim, M.~Kaminsky, and D.~G. Andersen.
\newblock Cicada: Dependably fast multi-core in-memory transactions.
\newblock In {\em {SIGMOD} Conference}, pages 21--35, 2017.

\bibitem{LinC0OTW16}
Q.~Lin, P.~Chang, G.~Chen, B.~C. Ooi, K.~Tan, and Z.~Wang.
\newblock Towards a non-2{PC} transaction management in distributed database
  systems.
\newblock In {\em SIGMOD Conference}, pages 1659--1674, 2016.

\bibitem{MahmoudANAA14}
H.~A. Mahmoud, V.~Arora, F.~Nawab, D.~Agrawal, and A.~{El Abbadi}.
\newblock {MaaT}: Effective and scalable coordination of distributed
  transactions in the cloud.
\newblock {\em {PVLDB}}, 7(5):329--340, 2014.

\bibitem{MalviyaWMS14}
N.~Malviya, A.~Weisberg, S.~Madden, and M.~Stonebraker.
\newblock Rethinking main memory {OLTP} recovery.
\newblock In {\em ICDE}, pages 604--615, 2014.

\bibitem{MaoJM08}
Y.~Mao, F.~P. Junqueira, and K.~Marzullo.
\newblock Mencius: Building efficient replicated state machine for wans.
\newblock In {\em OSDI}, pages 369--384, 2008.

\bibitem{MaoKM12}
Y.~Mao, E.~Kohler, and R.~T. Morris.
\newblock Cache craftiness for fast multicore key-value storage.
\newblock In {\em EuroSys}, pages 183--196, 2012.

\bibitem{DB2}
D.~McInnis.
\newblock The basics of {DB2} log shipping.
\newblock
  \url{https://www.ibm.com/developerworks/data/library/techarticle/0304mcinnis/0304mcinnis.html},
  2003.

\bibitem{SQLServer}
Microsoft.
\newblock About log shipping ({SQL Server}).
\newblock \url{https://msdn.microsoft.com/en- us/library/ms187103.aspx}, 2016.

\bibitem{MohanLO86}
C.~Mohan, B.~G. Lindsay, and R.~Obermarck.
\newblock Transaction management in the {R*} distributed database management
  system.
\newblock {\em {ACM} Trans. Database Syst.}, 11(4):378--396, 1986.

\bibitem{MuCZLL14}
S.~Mu, Y.~Cui, Y.~Zhang, W.~Lloyd, and J.~Li.
\newblock Extracting more concurrency from distributed transactions.
\newblock In {\em OSDI}, pages 479--494, 2014.

\bibitem{MySQL}
{MySQL}.
\newblock {MySQL} 8.0 reference manual,.
\newblock \url{https://dev.mysql.com/doc/refman/8.0/en/replication.html}, 2019.

\bibitem{NarulaCKM14}
N.~Narula, C.~Cutler, E.~Kohler, and R.~Morris.
\newblock Phase reconciliation for contended in-memory transactions.
\newblock In {\em OSDI}, pages 511--524, 2014.

\bibitem{NeumannMK15}
T.~Neumann, T.~M{\"{u}}hlbauer, and A.~Kemper.
\newblock Fast serializable multi-version concurrency control for main-memory
  database systems.
\newblock In {\em SIGMOD Conference}, pages 677--689, 2015.

\bibitem{OngaroO14}
D.~Ongaro and J.~K. Ousterhout.
\newblock In search of an understandable consensus algorithm.
\newblock In {\em ATC}, pages 305--319, 2014.

\bibitem{PlattnerA04}
C.~Plattner and G.~Alonso.
\newblock Ganymed: Scalable replication for transactional web applications.
\newblock In {\em Middleware}, pages 155--174, 2004.

\bibitem{PlattnerAO08}
C.~Plattner, G.~Alonso, and M.~T. {\"{O}}zsu.
\newblock Extending {DBMS}s with satellite databases.
\newblock {\em {VLDB} J.}, 17(4):657--682, 2008.

\bibitem{PostgreSQL}
PostgreSQL.
\newblock {PostgreSQL} 9.6.13 documentation.
\newblock \url{https://www.postgresql.org/docs/9.6/static/warm-standby.html},
  2019.

\bibitem{QinGB17}
D.~Qin, A.~Goel, and A.~D. Brown.
\newblock Scalable replay-based replication for fast databases.
\newblock {\em {PVLDB}}, 10(13):2025--2036, 2017.

\bibitem{SerafiniTEPAS16}
M.~Serafini, R.~Taft, A.~J. Elmore, A.~Pavlo, A.~Aboulnaga, and M.~Stonebraker.
\newblock Clay: Fine-grained adaptive partitioning for general database
  schemas.
\newblock {\em {PVLDB}}, 10(4):445--456, 2016.

\bibitem{ShuteVSHWROLMECRSA13}
J.~Shute, R.~Vingralek, B.~Samwel, B.~Handy, C.~Whipkey, E.~Rollins, M.~Oancea,
  K.~Littlefield, D.~Menestrina, S.~Ellner, J.~Cieslewicz, I.~Rae,
  T.~Stancescu, and H.~Apte.
\newblock {F1:} {A} distributed {SQL} database that scales.
\newblock {\em {PVLDB}}, 6(11):1068--1079, 2013.

\bibitem{StonebrakerMAHHH07}
M.~Stonebraker, S.~Madden, D.~J. Abadi, S.~Harizopoulos, N.~Hachem, and
  P.~Helland.
\newblock The end of an architectural era (it's time for a complete rewrite).
\newblock In {\em VLDB}, pages 1150--1160, 2007.

\bibitem{Thomas79}
R.~H. Thomas.
\newblock A majority consensus approach to concurrency control for multiple
  copy databases.
\newblock {\em TODS}, 4(2):180--209, 1979.

\bibitem{ThomsonA10}
A.~Thomson and D.~J. Abadi.
\newblock The case for determinism in database systems.
\newblock {\em {PVLDB}}, 3(1):70--80, 2010.

\bibitem{ThomsonDWRSA12}
A.~Thomson, T.~Diamond, S.~Weng, K.~Ren, P.~Shao, and D.~J. Abadi.
\newblock Calvin: fast distributed transactions for partitioned database
  systems.
\newblock In {\em SIGMOD Conference}, pages 1--12, 2012.

\bibitem{TuZKLM13}
S.~Tu, W.~Zheng, E.~Kohler, B.~Liskov, and S.~Madden.
\newblock Speedy transactions in multicore in-memory databases.
\newblock In {\em SOSP}, pages 18--32, 2013.

\bibitem{VandiverBLM07}
B.~Vandiver, H.~Balakrishnan, B.~Liskov, and S.~Madden.
\newblock Tolerating byzantine faults in transaction processing systems using
  commit barrier scheduling.
\newblock In {\em SOSP}, pages 59--72, 2007.

\bibitem{WangK16}
T.~Wang and H.~Kimura.
\newblock Mostly-optimistic concurrency control for highly contended dynamic
  workloads on a thousand cores.
\newblock {\em {PVLDB}}, 10(2):49--60, 2016.

\bibitem{WeiSCCC15}
X.~Wei, J.~Shi, Y.~Chen, R.~Chen, and H.~Chen.
\newblock Fast in-memory transaction processing using {RDMA} and {HTM}.
\newblock In {\em SOSP}, pages 87--104, 2015.

\bibitem{YuPSD16}
X.~Yu, A.~Pavlo, D.~Sanchez, and S.~Devadas.
\newblock {TicToc}: Time traveling optimistic concurrency control.
\newblock In {\em SIGMOD Conference}, pages 1629--1642, 2016.

\bibitem{ZhengTKL14}
W.~Zheng, S.~Tu, E.~Kohler, and B.~Liskov.
\newblock Fast databases with fast durability and recovery through multicore
  parallelism.
\newblock In {\em OSDI}, pages 465--477, 2014.

\end{thebibliography}

\end{document}